\documentclass[12pt,twoside,a4paper]{article}
\usepackage{eurosym}
\usepackage{amsmath}
\usepackage{multirow}
\usepackage{graphicx}
\usepackage{booktabs}
\usepackage[section]{placeins}
\usepackage[top=1.5in,bottom=1in,right=1in,left=1in,headheight=15pt]{geometry}
\usepackage{setspace}
\usepackage{rotating}
\usepackage{textcomp}
\usepackage{amssymb}
\usepackage{amsmath}
\usepackage{siunitx}
\usepackage{etoolbox}
\usepackage{amsmath,amssymb}
\usepackage{float}
\usepackage{xcolor}
\usepackage{hyperref}
\usepackage{booktabs,caption,fixltx2e}
\usepackage[flushleft]{threeparttable}
\usepackage{tabularx}
\usepackage{booktabs,caption,fixltx2e}
\usepackage[flushleft]{threeparttable}
\usepackage{tabularx}
\usepackage{enumitem}
\usepackage{caption}
\usepackage{nameref}
\usepackage{lscape}
\usepackage{fancyhdr}

\hypersetup{
    colorlinks, linkcolor={dark-red},
    citecolor={blue}, urlcolor={blue}, linkcolor={blue}
}
\begin{document}

\title{Classifying Patents Based on their Semantic Content\thanks{%
Addresses - Bergeaud: London School of Economics; Potiron: Faculty of Business and Commerce, Keio University, Tokyo, Japan;
Raimbault: UMR CNRS 8504 G{\'e}ographie-cit{\'e}s, Universit{\'e} Paris VII, Paris, France and UMR-T 9403 IFSTTAR LVMT, Ecole Nationale des Ponts et Chauss{\'e}es, Champs-sur-Marne, France.} \vspace{5mm} }
\author{Antonin Bergeaud \ \ \ \ \ Yoann Potiron
 \ \ \ \ \ Juste Raimbault \bigskip}
\maketitle

\begin{abstract}
In this paper, we extend some usual techniques of classification resulting from a large-scale data-mining and network approach. This new technology, which in particular is designed to be suitable to big data, is used to construct an open consolidated database from raw data on 4 million patents taken from the US patent office from 1976 onward. To build the pattern network, not only do we look at each patent title, but we also examine their full abstract and extract the relevant keywords accordingly. We refer to this classification as \emph{semantic approach} in contrast with the more common \emph{technological approach} which consists in taking the topology when considering US Patent office technological classes. Moreover, we document that both approaches have highly different topological measures and strong statistical evidence that they feature a different model. This suggests that our method is a useful tool to extract endogenous information.    
\end{abstract}

\pagebreak

\section{Introduction}
Innovation and technological change have been described by many scholars as the main drivers of economic growth as in \cite{aghionhowitt1992} and \cite{romer1990}. \cite{RePEc:nbr:nberwo:3301} advertised the use of patents as an economic indicator and as a good proxy for innovation. Subsequently, the easier availability of comprehensive databases on patent details and the increasing number of studies allowing a more efficient use of these data (e.g.~\cite{Hall2001}) have opened the way to a very wide range of analysis. Most of the statistics derived from the patent databases relied on a few key features: the identity of the inventor, the type and identity of the rights owner, the citations made by the patent to prior art and the technological classes assigned by the patent office post patent's content review. Combining this information is particularly relevant when trying to capture the diffusion of knowledge and the interaction between technological fields as studied in \cite{Youn:2015fk}. With methods such as citation dynamics modeling discussed in~\cite{2013arXiv1310.8220N} or co-authorship networks analysis in~\cite{2014arXiv1402.7268S}, a large body of the literature such as~\cite{sorenson2006complexity} or~\cite{kay2014patent} has studied patents citation network to understand processes driving technological innovation, diffusion and the birth of technological clusters. Finally, \cite{bruck2016recognition} look at the dynamics of citations from different classes to show that the laser/ink-jet printer technology resulted from the recombination of two different existing technologies. 

Consequently, technological classification combined with other features of patents can be a valuable tool for researchers interested in studying technologies throughout history and to predict future innovations by looking at past knowledge and interaction across sectors and technologies. But it is also crucial for firms that face an ever changing demand structure and need to anticipate future technological trends and convergence (see, e.g., \cite{curran2011patent}) to adapt to the resulting increase in competition discussed in~\cite{Katz1996remarks} and to maintain market share. Curiously, and in spite of the large number of studies that analyze interactions across technologies~\cite{Furman2011shoulders}, little is known about the underlying ``innovation network'' (e.g. ~\cite{AAKnetwork2016}). 

In this monograph, we propose an alternative classification based on semantic network analysis from patent abstracts and explore the new information emerging from it. In contrast with the regular technological classification which results from the choice of the patent reviewer, semantic classification is carried automatically based on the content of the patent abstract. Although patent officers are experts in their fields, the relevance of the existing classification is limited by the fact that it is based on the state of technology at the time the patent was granted and cannot anticipate the birth of new fields.\footnote{To correct for this, the USPTO regularly make changes in its classification in order to adapt to technological change (for example, the ``nanotechnology'' class (977) was established in 2004 and  retroactively to all relevant previously granted patents).} In contrast we don't face this issue with the semantic approach. The semantic links can be clues of one technology taking inspiration from another and  good predictors of future technology convergence (e.g.~\cite{preschitschek2013} study semantic similarities from the whole text of 326 US-patents on \textit{phytosterols} and show that semantic analysis have a good predicting power of future technology convergence). One can for instance consider the case of the word \textit{optic}. Until more recently, this word was often associated with technologies such as photography or eye surgery, while it is now almost exclusively used in a context of semi-transistor design and electro-optic. This semantic shift did not happen by chance but contains information on the fact that modern electronic extensively uses technologies that were initially developed in optic. 

Previous research has already proposed to use semantic networks to study technological domains and detect novelty. ~\cite{yoon2004text} was one of the first to enhance this approach with the idea of visualizing keywords network illustrated on a small technological domain. The same approach can be used to help companies identifying the state of the art in their field and avoid patent infringement as in \cite{park2014semantic} and \cite{yoon2011detecting}. More closely related to our methodology,~\cite{gerken2012new} develop a method based on patent semantic analysis of patent to vindicate the view that this approach outperform others in the monitoring of technology and in the identification of novelty innovation. Semantic analysis has already proven its efficiency in various fields, such as in technology studies (e.g.~\cite{choi2014patent} and~\cite{fattori2003text}) and in political science (e.g.~\cite{2015arXiv151003797G}).

Building on such previous research, we make several contributions by fulfilling some shortcomings of existing studies, such as for example the use of frequency-selected single keywords. First of all, we develop and implement a novel fully-automatized methodology to classify patents according to their semantic abstract content, which is to the best of our knowledge the first of its type. This includes the following refinements for which details can be found in Section \ref{keywords}: (i) use of multi-stems as potential keywords; (ii) filtering of keywords based on a second-order (co-occurrences) relevance measure and on an external independent measure (technological dispersion); (iii) multi-objective optimization of semantic network modularity and size. The use of all this techniques in the context of semantic classification is new and essential from a practical perspective. 

Furthermore, most of the existing studies rely on a subsample of patent data, whereas we implement it on the full US Patent database from 1976 to 2013. This way, a general structure of technological innovation can be studied. We draw from this application promising qualitative stylized facts, such as a qualitative regime shift around the end of the 1990s, and a significant improvement of citation modularity for the semantic classification when comparing to the technological classification. These thematic conclusions validate our method as a useful tool to extract endogenous information, in a complementary way to the technological classification.

Finally, the statistical model introduced in Section \ref{statisticalmodel} seems to indicate that patents tend to cite more similar patents in the semantic network  when fitted to data. In particular, this propensity is shown to be significantly bigger than the corresponding propensity for technological classes, and this seems to be consistent over time. On the account of this information, we believe that patent officers could benefit very much from looking at the semantic network when considering potential citation candidates of a patent in review.

The paper is organized as follows. Section \ref{data} presents the patent data, the existing classification and provide details about the data collection process. Section \ref{keywords} explains the construction of the semantic classes. Section \ref{result} tests their relevance by providing exploratory results. Finally, section \ref{discussion} discusses potential further developments and conclude. More details, including robustness checking, figures and technical derivations can be found in~\nameref{sectionSI}.

\section{Background \label{data}}

In our analysis, we will consider all utility patents granted in the United States Patent and Trademark Office (USPTO) from 1976 to 2013. A clearer definition of utility patent is given in~\nameref{sectionSI}. Also, additional information on how to correctly exploit patent data can be found in \cite{Hall2001} and \cite{lerner2015use}.

\subsection{An existing classification: the USPC system}

Each USPTO patent is associated with a non-empty set of technological classes and subclasses. There are currently around 440 classes and over 150,000 subclasses constituting the United State Patent Classification (USPC) system. While a technological class corresponds to the technological field covered by the patent, a subclass stands for a specific technology or method used in this invention. A patent can have multiple technological classes, on average in our data a patent has 1.8 different classes and 3.9 pairs of class/subclass. At this stage, two features of this system are worth mentioning: (i) classes and subclasses are not chosen by the inventors of the patent but by the examiner during the granting process based on the content of the patent; (ii) the classification has evolved in time and continues to change in order to adapt to new technologies by creating or editing classes. When a change occurs, the USPTO reviews all the previous patents so as to create a consistent classification.

\subsection{A bibliographical network between patents: citations \label{sub:citation}}

As with scientific publications, patents must give reference to all the previous patents which correspond to related prior art. They therefore indicate the past knowledge which relates to the patented invention. Yet, contrary to scientific citations, they also have an important legal role as they are used to delimit the scope of the property rights awarded by the patent. One can consult~\cite{oecdpatentmanual} for more details about this. Failing to refer to prior art can lead to the invalidation of the patent (e.g.~\cite{martin2015}). Another crucial difference is that the majority of the citations are actually chosen by the  examiners and not by the inventors themselves. 
From the USPTO, we gather information of all citations made by each patent (backward citations) and all citations received by each patent as of the end of 2013 (forward citations). We can thus build a complete network of citations that we will use later on in the analysis.

Turning to the structure of the lag between the citing and the cited patent in terms of application date, we see that the mean of this lag is 8.5 years and the median is 7 years. This distribution is highly skewed, the $95^{th}$ percentile is 21 years. We also report 164,000 citations with a negative time lag. This is due to the fact that some citations can be added during the examination process and some patents require more time to be granted than others.

In what follows, we choose to restrict attention to pairs of citations with a lag no larger than 5 years. We impose this restriction for two reasons. First, the number of citations received peaks 4-5 years after application. Second, the structure of the citation lag is necessarily biased by the truncation of our sample: the more recent patents mechanically receive less citations than the older ones. As we are restricting to citations received no later than 5 years after the application date, this effect will only affect patents with an application date after 2007.

\subsection{Data collection and basic description}
Each patent contains an abstract and a core text which describe the invention.\footnote{To see what a patent looks like in practice, one can refer to the USPTO patent full-text database \url{http://patft.uspto.gov/netahtml/PTO/index.html} or to Google patent which publishes USPTO patents in $pdf$ form at \url{https://patents.google.com}.} Although including the full core texts would be natural and probably very useful in a systematic text-mining approach as done in~\cite{tseng2007text}, they are too long to be included and thus we consider only the abstracts for the analysis. Indeed, the semantic analysis counts more than 4 million patents, with corresponding abstracts with an average length of 120.8 words (and a standard deviation of $62.4$), a size that is already challenging in terms of computational burden and data size. In addition, abstracts are aimed at synthesizing purpose and content of patents and must therefore be a relevant object of study (see~\cite{Adams2010text}). The USPTO defines a guidance stating that an abstract should be ``a summary of the disclosure as contained in the description, the claims, and any drawings; the summary shall indicate the technical field to which the invention pertains and shall be drafted in a way which allows the clear understanding of the technical problem, the gist of the solution of that problem through the invention, and the principal use or uses of the invention'' (PCT Rule 8). 

We construct from raw data a unified database. Data is collected from USPTO patent redbook bulk downloads, that provides as raw data (specific \texttt{dat} or \texttt{xml} formats) full patent information, starting from 1976. Detailed procedure of data collection, parsing and consolidation are available in~\nameref{sectionSI}. The latest dump of the database in \texttt{Mongodb} format is available at 
\url{http://dx.doi.org/10.7910/DVN/BW3ACK} 
Collection and homogenization of the database into a directly usable database with basic information and abstracts was an important task as USPTO raw data formats are involved and change frequently.

We count 4,666,365 utility patents with an abstract granted from 1976 to 2013.\footnote{A very small number of patents have a missing abstract, these are patents that have been withdrawn and we do not consider them in the analysis.} The number of patents granted each year increases from around 70,000 in 1976 to about 278,000 in 2013. When distributed by the year of application, the picture is slightly different. The number of patents steadily increase from 1976 to 2000 and remains constant around 200,000 per year from 2000 to 2007. Restricting our sample to patent with application date ranging from 1976 to 2007, we are left with 3,949,615 patents. These patents cite 38,756,292 other patents with the empirical lag distribution that has been extensively analyzed in~\cite{Hall2001}. Conditioned on being cited at least once, a patent receives on average 13.5 citations within a five-year window. 270,877 patents receive no citation during the next five years following application, 10\% of patents receive only one citation and 1\% of them receive more than 100 citations. A within class citation is defined as a citation between two patents sharing at least one common technological class. Following this definition, 84\% of the citations are within class citations. 14\% of the citations are between two patents that share the exact same set of technological classes.

\subsection{Towards a Complementary Classification}

Potentialities of text-mining techniques as an alternative way to analyze and classify patents are documented in~\cite{tseng2007text}. The author's main argument, in support of an automatic classification tool for patent, is to reduce the considerable amount of human effort needed to classify all the applications. The work conducted in the field of natural language processing and/or text analysis has been developed in order to improve search performance in patent databases, build technology map or investigate the potential infringement risks prior to developing a new technology (see~\cite{abbas2014literature} for a review). Text-mining of patent documents is also widely used as a tool to build networks which carry additional information to the simplistic bibliographic connections model as argued in~\cite{yoon2004text}. As far as the authors know, the use of text-mining as a way to build a global classification of patents remains however largely unexplored. One notable exception can be found in~\cite{preschitschek2013} where semantic-based classification is shown to outperform the standard classification in predicting the convergence of technologies even in small samples. Semantic analysis reveals itself to be more flexible and more quickly adaptable to the apparition of new clusters of technologies. Indeed, as argued in~\cite{preschitschek2013}, before two distinct technologies start to clearly converge, one should expect similar words to be used in patents from both technologies.

Finally, a semantic classification where patents are gathered based on the fact that they share similar significant keywords has the advantage of including a network feature that cannot be found in the USPC case, namely that each patent is associated with a vector of probability to belong to each of the semantic classes (more details on this feature can be found in Section \ref{characteristics}). Using co-occurrence of keywords, it is then possible to construct a network of patents and to study the influence of some key topological features.

\section{Semantic Classification Construction \label{keywords}}

In this section, we describe methods and empirical analysis leading to the construction of semantic network and the corresponding classification. 

\subsection{Keywords extraction}
\label{keywordsextraction}

Let $\mathcal{P}$ be the set of patents, we first assign to a patent $p\in \mathcal{P}$ a set of potentially significant keywords $K(p)$ from its text ${\mathcal{A}}(p)$ (that corresponds to the concatenation of its own title and abstract).
$K(p)$ are extracted through a similar procedure as the one detailed in~\cite{chavalarias2013phylomemetic}: 
\begin{enumerate}
\item Text parsing and Tokenization: we transform raw texts into a set of words and sentences, reading it (parsing) and splitting it into elementary entities (words organized in sentences).

\item Part-of-speech tagging: attribution of a grammatical function to each of the tokens defined previously.

\item Stem extraction: families of words are generally derived from a unique root called stem (for example \texttt{compute}, \texttt{computer}, \texttt{computation} all yield the same stem \texttt{comput}) that we extract from tokens. At this point the abstract text is reduced to a set of stems and their grammatical functions.
\item Multi-stems construction: these are the basic semantic units used in further analysis. They are constructed as groups of successive stems in a sentence which satisfies a simple grammatical function rule. The length of the group is between 1 and 3 and its elements are either nouns, attributive verbs or adjectives. We choose to extract the semantics from such nominal groups in view of the technical nature of texts, which is not likely to contain subtle nuances in combinations of verbs and nominal groups.
\end{enumerate}

Text processing operations are implemented in \texttt{python} in order to use built-in functions \texttt{nltk} library~\cite{nltk} for most of above operations. This library supports most of state-of-the-art natural language processing operations.\footnote{Source code is openly available on the repository of the project: \url{https://github.com/JusteRaimbault/PatentsMining} 
}

\subsection{Keywords relevance estimation}
\label{keywords_est}
\paragraph*{Relevance definition}

Following the heuristic in~\cite{chavalarias2013phylomemetic}, we estimate relevance score in order to filter multi-stem. The choice of the total number of keywords to be extracted $K_w$ is important, too small a value would yield similar network structures but including less information whereas very large values tend to include too many irrelevant keywords. We choose to set this parameter to $K_w = 100,000$. We first consider the filtration of $k\cdot K_w$ (with $k=4$) to keep a large set of potential keywords but still have a reasonable number of co-occurrences to be computed. This is done on the \emph{unithood} $u_i$, defined for keyword $i$ as $u_i = f_i\cdot \log{(1 + l_i)}$ where $f_i$ is the multi-stem's number of apparitions over the whole corpus and $l_i$ its length in words. A second filtration of $K_w$ keywords is done on the \emph{termhood} $t_i$. The latter is computed as a chi-squared score on the distribution of the stem's co-occurrences and then compared to a uniform distribution within the whole corpus. Intuitively, uniformly distributed terms will be identified as plain language and they are thus not relevant for the classification. More precisely, we compute the co-occurrence matrix $(M_{ij})$, where $M_{ij}$ is defined as the number of patents where stems $i$ and $j$ appear together. The \emph{termhood} score $t_i$ is defined as

\[
t_i = \sum_{j\neq i}\frac{\left( M_{ij} - \sum_{k}M_{ik} \sum_{k} M_{jk}\right)^2}{\sum_{k}M_{ik} \sum_{k} M_{jk}}.
\]

\paragraph*{Moving window estimation}
The previous scores are estimated on a moving window with fixed time length following the idea that the present relevance is given by the most recent context and thus that the influence vanishes when going further into the past. Consequently, the co-occurrence matrix is chosen to be constructed at year $t$ restricting to patent which applied during the time window $\big[ t - T_0 ; t \big]$. Note that the causal property of the window is crucial as the future cannot play any role in the current state of keywords and patents. This way, we will obtain semantic classes which are exploitable on a $T_0$ time span. For example, this enables us to compute the modularity of classes in the citation network as in section \ref{citationmodularity}. In the following, we take $T_0 = 4$ (which corresponds to a five year window) consistently with the choice of maximum time lag for citations made in Section \ref{sub:citation}. Accordingly, the sensitivity analysis for $T_0=2$ can be found in Appendix~\nameref{app:sensitivity}.

\subsection{Construction of the semantic network}
\label{construction}

\begin{figure}
\centering
\includegraphics[width=0.45\textwidth,height=0.25\textheight]{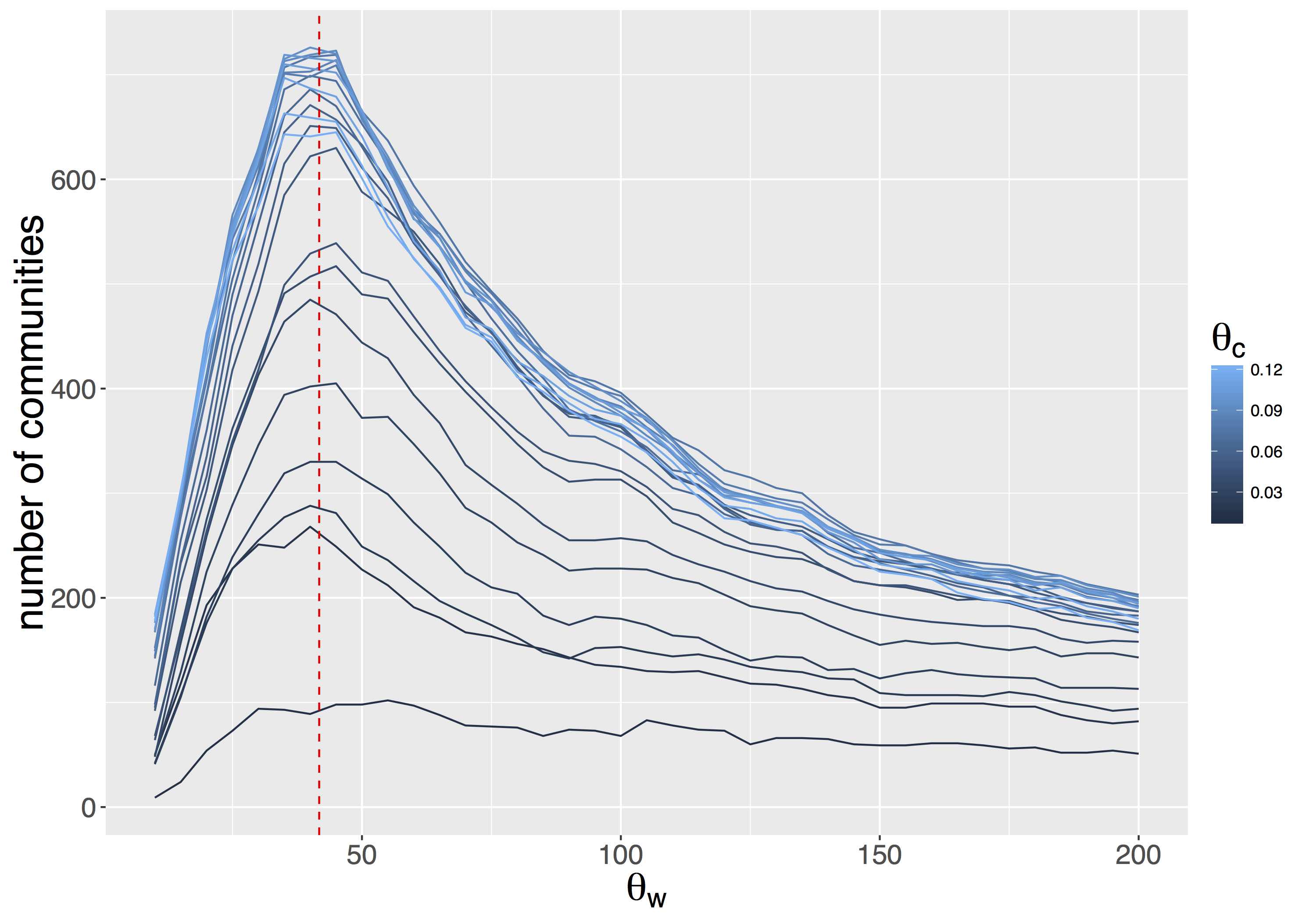}
\includegraphics[width=0.45\textwidth,height=0.25\textheight]{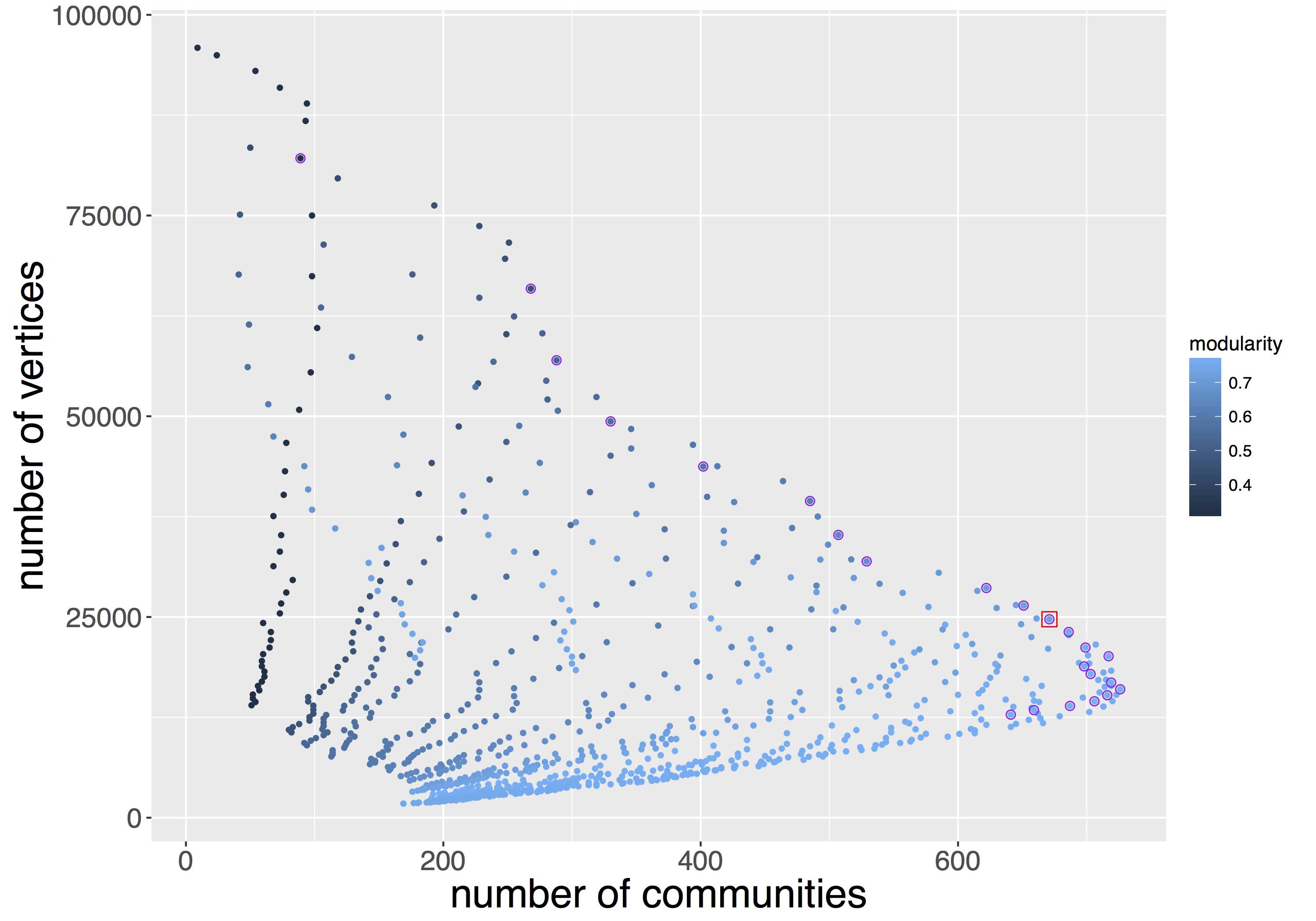}
\caption{\textbf{Sensitivity analysis of network community structure to filtering parameters.} We consider a specific window 2000-2004 and the obtained plots are typical. \textit{(Left panel)} We plot the number of communities as a function of $\theta_w$ for different $\theta_c$ values. The maximum is roughly stable across $\theta_c$ (dashed red line). \textit{(Right panel)} To choose $\theta_c$, we do a Pareto optimization on communities and network size: the compromise point (red overline) on the Pareto front (purple overline: possible choices after having fixed $\theta_w^{(0)}$; blue level gives modularity) corresponds to $\theta_c = 0.06$.}
\label{fig:networksensitivity}
\end{figure}

We keep the set of most relevant keywords $\mathcal{K}_W$ and obtain their co-occurrence matrix as defined in Section \ref{keywords_est}. This matrix can be directly interpreted as the weighted adjacency matrix of the semantic network. At this stage, the topology of raw networks does not allow the extraction of clear communities. This is partly due to the presence of hubs that correspond to frequent terms common to many fields (e.g. \texttt{method}, \texttt{apparat}) which are wrongly filtered as relevant. We therefore introduce an additional measure to correct the network topology: the concentration of keywords across technological classes, defined as: 

$$c_{tech}(s) = \displaystyle \sum_{j=1}^{N^{(tec)}} \frac{k_j(s)^2}{ \left(\sum_i k_i(s)\right)^2},$$  

where $k_j(s)$ is the number of occurrences of the $s$th keyword in each of the $j$th technological class taken from one of the $N^{(tec)}$ USPC classes. The higher $c_{tech}$, the more specific to a technological class the node is. For example, the terms \texttt{semiconductor} is widely used in electronics and does not contain any significant information in this field. We use a threshold parameter and keep nodes with $c_{tech}(s) > \theta_c$. In a similar manner, edge with low weights correspond to rare co-occurrences and are considered as noise: we filter edges with weight lower than a threshold $\theta_w$, following the rationale that two keywords are not linked ``by chance'' if they appear simultaneously a minimal number of time. To control for size effect, we normalize by taking $\theta_w = \theta_w^{(0)}\cdot N_P$ where $N_P$ is the number of patents in the corpus ($N_P = \left|\mathcal{P} \right|$). Communities are then extracted using a standard modularity maximization procedure as described in~\cite{clauset2004finding} to which we add the two constraints captured by $\theta_w$ and $\theta_c$, namely that edges must have a weight greater than $\theta_w$ and nodes a concentration greater than $\theta_c$. At this stage, both parameters $\theta_c$ and $\theta_w^{(0)}$ are unconstrained and their choice is not straightforward. Indeed, many optimization objectives are possible, such as the modularity, network size or number of communities. We find that modularity is maximized at a roughly stable value of $\theta_w$ across different $\theta_c$ for each year, corresponding to a stable $\theta_w^{(0)}$ across years, which leads us to choose $\theta_w^{(0)} = 4.1\cdot 10^{-5}$. Then for the choice of $\theta_c$, different candidates points lie on a Pareto front for the bi-objective optimization on number of communities and network size, among which we take $\theta_c = 0.06$ (see Fig.~\ref{fig:networksensitivity}).

\begin{figure}
\centering
\includegraphics[width=\textwidth]{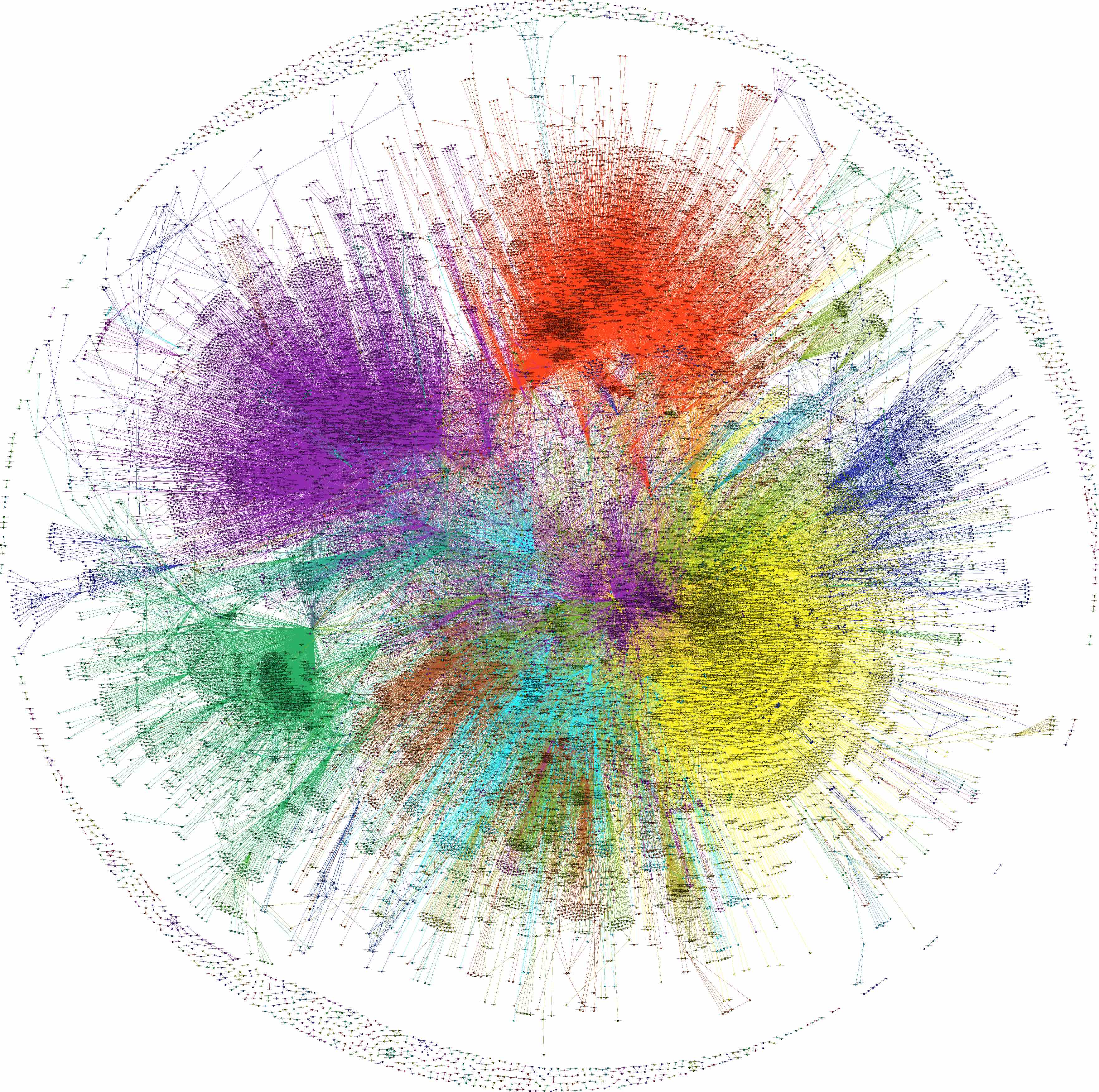}
\caption{\textbf{An example of semantic network visualization.} We show the network obtained for the window 2000-2004, with parameters $\theta_c = 0.06$ and $\theta_w = \theta_w^{(0)}\cdot N_P = 4.5e^{-5} \cdot 9.1e^{5}$. The corresponding file in a vector format (\texttt{.svg}), that can be zoomed and explored, is available at \texttt{http://37.187.242.99/files/public/network.svg}.}
\label{fig:rawnetwork}
\end{figure}

\subsection{Characteristics of Semantic Classes}
\label{characteristics}
\begin{figure}
\centering
\includegraphics[width=0.9\textwidth]{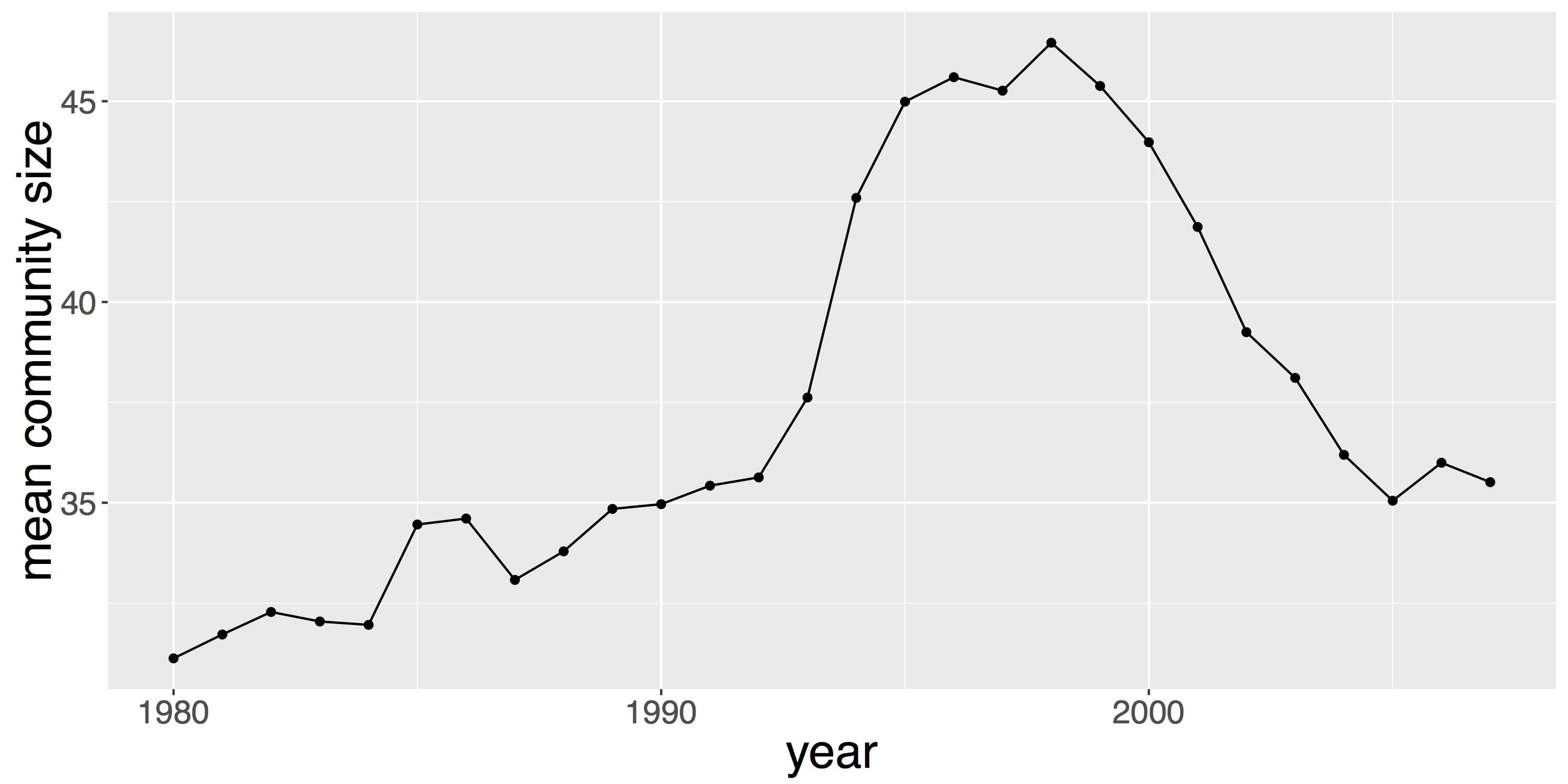}
\caption{\textbf{This figure plots the average number of keywords by semantic class for each time window $\left[t-4; t\right]$} from $t=1980$ to $t=2007$.}
\label{fig:mean_K}
\end{figure}

For each year $t$, we define as $N^{(sem)}_t$ the number of semantic classes which have been computed by clustering keywords from patents appeared during the period $\big[ t-T_0, t \big]$ (we recall that we have chosen $T_0=4$). Each semantic class $k =  1, \cdots, N^{(sem)}_t$ is characterized by a set of keywords $K(k,t)$ which is a subset of $\mathcal{K}_W$ selected as described in Section \ref{keywordsextraction} to Section \ref{construction}. The cardinal of $K(k, t)$ distribution across each semantic class $k$ is highly skewed with a few semantic classes containing over $1,000$ keywords, most of them with roughly the same number of keywords. In contrast, there are also many semantic classes with only two keywords. There are around 30 keywords by semantic class on average and the median is 2 for any $t$. Fig.~\ref{fig:mean_K} shows that the average number of keywords is relatively stable from 1976 to 1992 and then picks around 1996 prior to going down.

\paragraph*{Title of semantic classes}
USPC technological classes are defined by a title and a highly accurate definition which help retrieve patents easily. The title can be a single word (e.g.: class 101: ``Printing'') or more complex (e.g.: class 218: ``High-voltage switches with arc preventing or extinguishing devices''). As our goal is to release a comprehensive database in which each patent is associated with a set of semantic classes, it is necessary to give an insight on what these classes represent by associating a short description or a title as in~\cite{tseng2007text}. In our case, such description is taken as a subset of keywords taken from $K(k,t)$. For the vast majority of semantic classes that have less than 5 keywords, we decide to keep all of theses keywords as a description. For the remaining classes which feature around 50 keywords on average, we rely on the topological properties of the semantic network.~\cite{yang2000improving} suggest to retain only the most frequently used terms in $K(k,t)$. Another possibility is to select 5 keywords based on their network centrality with the idea that very central keywords are the best candidates to describe the overall idea captured by a community. For example, the largest semantic class in 2003-2007 is characterized by the keywords: \texttt{Support Packet; Tree Network; Network Wide; Voic Stream; Code Symbol Reader}.

\begin{figure}
\centering
\hspace*{-2.5cm}
\includegraphics[width=1.2\textwidth]{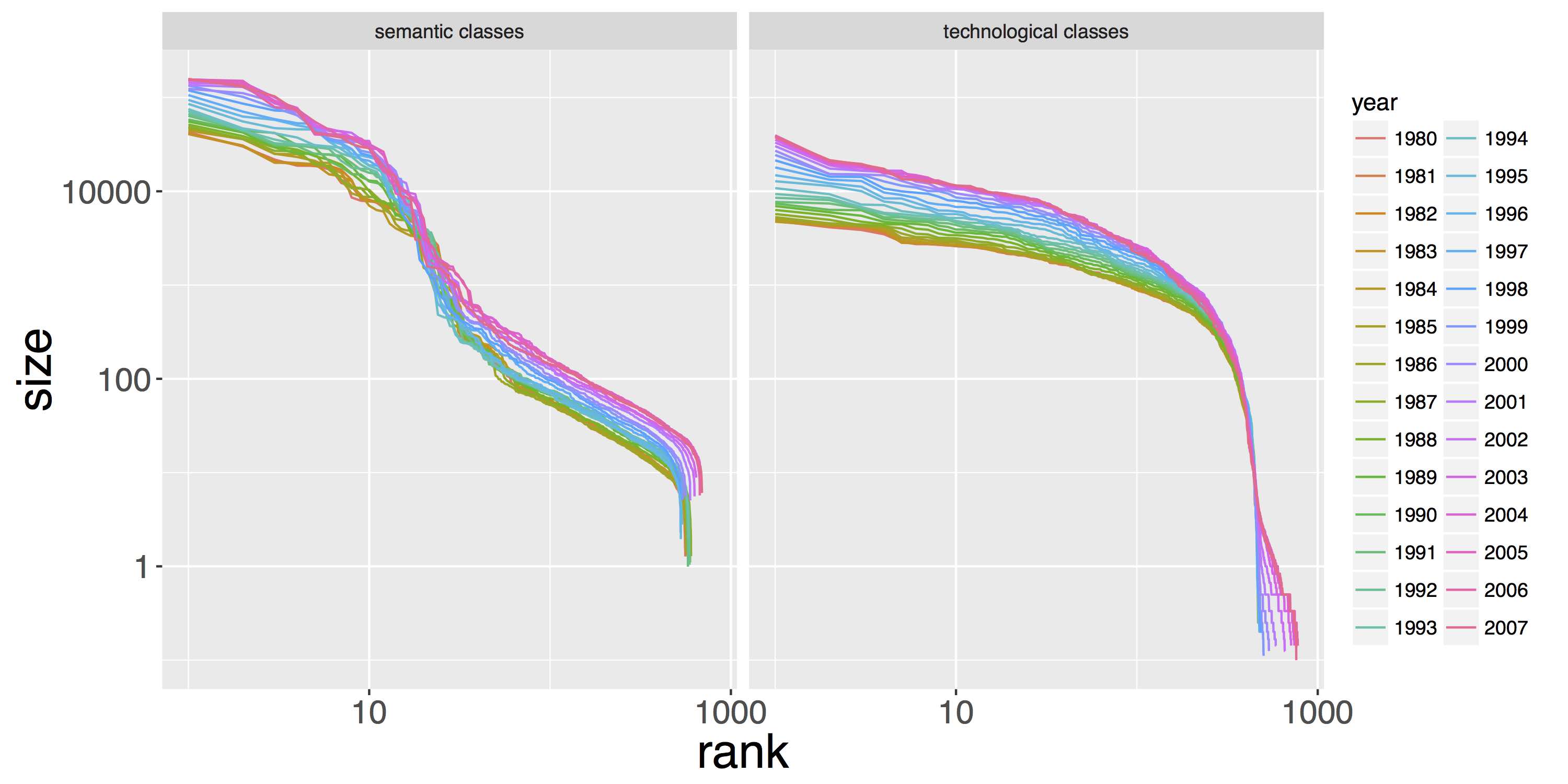}
\caption{\textbf{Sizes of classes.} Yearly from $t = 1980$ to $t =2007$, we plot the size of semantic classes (left-side) and technological classes (right-side) for the corresponding time window $[t-4, t]$, 
from the biggest to the smallest. The formal definition of size can be found in Section \ref{characteristics}. Each color corresponds to one specific year. Yearly semantic classes and technological classes present a similar hierarchical structure which confirms the comparability of the two classifications. This feature is crucial for the statistical analysis in Section \ref{statisticalmodel}. Over time, curves are translated and levels of hierarchy stays roughly constant.}
\label{fig:class-sizes}
\end{figure}

\paragraph*{Size of technological and semantic classes} 
We consider a specific window of observations (for example 2000-2004), and we define $Z$ the number of patents which appeared during that time window. For each patent $i=1, \cdots, Z$ we associate a vector of probability where each component $p_{ij}^{(sem)} \in \big[ 0,1 \big]$, with  $j = 1, \cdots, N{(sem)}$ and where\footnote{When there is no room for confusion, we drop the subscript $t$ in $N_t^{(sem)}$.} 
$$\displaystyle \sum_{j=1}^{N^{(sem)}} p_{ij}^{(sem)} = 1.$$ On average across all time windows, a patent is associated to 1.8 semantic classes with a positive probability. Next we define the size of a semantic class as $$S_j^{(sem)} = \displaystyle \sum_{i=1}^Z p_{ij}^{(sem)}.$$ 
Correspondingly, we aim to provide a consistent definition for technological classes. For that purpose, we follow the so-called ``fractional count'' method, which was introduced by the USPTO and consists in dividing equally the patents between all the classes they belong to. Formally, we define the number of technological classes as
$N^{(tec)}$  (which is not time dependent contrary to the semantic case) and for $j = 1, \cdots, N^{(tec)}$ the corresponding matrix of probability is defined as
\[
 p_{ij}^{(tec)} = \frac{B_{ij}}{\displaystyle \sum_{k=1}^{N^{(tec)}}{B_{ik}}},
\]
where  $B_{ij}$ equals $1$ if the $i$th patent belongs to the $j$th technological class and $0$ if not. When there is no room for confusion, we will drop the exponent part and write only $p_{ij}$ when referring to either the technological or semantic matrix. Empirically, we find that both classes exhibit a similar hierarchical structure in the sense of a power-law type of distribution of class sizes as shown in Fig.~\ref{fig:class-sizes}. This feature is important, it suggests that a classification based on the text content of patents has some separating power in the sense that it does not divide up all the patents in one or two communities. 

\subsection{Potential Refinements of the Method}

Our semantic classification method could be refined by combining it with other techniques such as Latent Dirichlet Allocation which is a widely used topic detection method (e.g.~\cite{blei2003latent}), already used on patent data as in~\cite{kaplan2015double} where it provides a measure of idea novelty and the counter-intuitive stylized facts that breakthrough invention are likely to come out of local search in a field rather than distant technological recombination. Using this approach should first help further evaluate the robustness of our qualitative conclusions (external validation). Also, depending on the level of orthogonality with our classification, it can potentially bring an additional feature to characterize patents, in the spirit of multi-modeling techniques where neighbor models are combined to take advantage of each point of view on a system.

Our use of network analysis can also be extended using newly developed techniques of hyper-network analysis. Indeed, patents and keywords can for example be nodes of a bipartite network, or patents be links of an hyper-network, in the sense of multiple layers with different classification links and citation links. \cite{iacovacci2015mesoscopic} provide a method to compare macroscopic structures of the different layers in a multilayer network that could be applied as a refinement of the overlap, modularity and statistical modeling studied in this paper. Furthermore, is has recently been shown that measures of multilayer network projections   induce a significant loss of information compared to the generalized corresponding measure~\cite{de2015ranking}, which confirms the relevance of such development that we left for further research.

\section{Results \label{result}}

In this section, we present some key features of our resulting semantic classification showing both complementary and differences with the technological classification. We first present several measures derived from this semantic classification at the patent level: Diversity, Originality, Generality (Section \ref{subsec:orig-gene}) and Overlapping (Section \ref{subsec:overlaps}). We then show that the two classifications show highly
different topological measures and strong statistical evidence that they feature a different model (Sections \ref{citationmodularity} and \ref{statisticalmodel}).


\subsection{Patent Level Measures}  \label{subsec:orig-gene}

Given a classification system (technological or semantic classes), and the associated probabilities $p_{ij}$ for each patent $i$ to belong to class $j$ (that were defined in Section \ref{characteristics}), one can define a patent-level diversity measure as one minus the Herfindhal concentration index on $p_{ij}$  by

\[
D_i^{(z)} = 1 - \sum_{j =1}^{N^{(z)}} {p_{ij}^2}, \text{ with } z \in \{tec, sem\}.
\]

\begin{figure}
\includegraphics[width=0.49\textwidth]{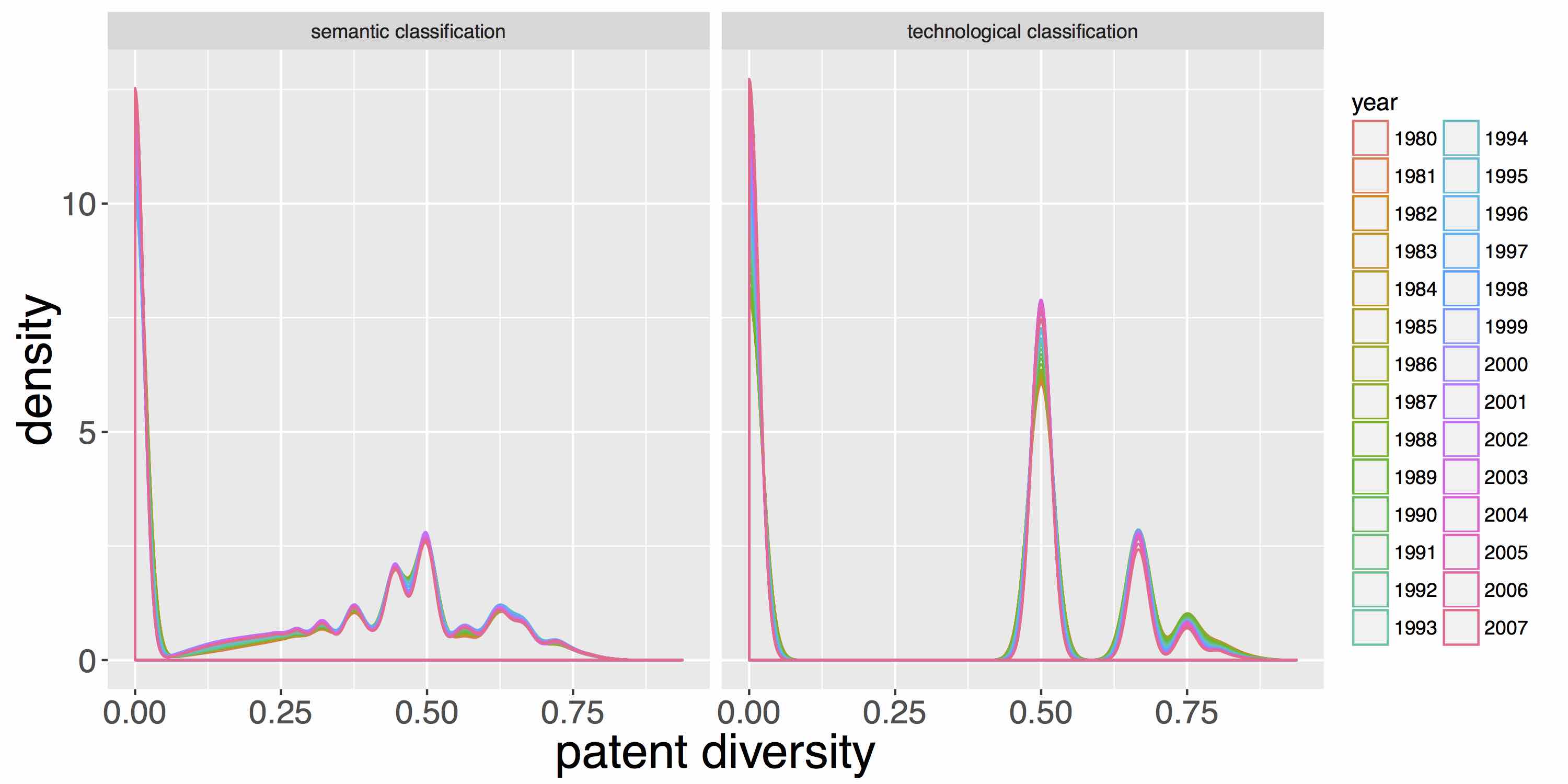}
\includegraphics[width=0.49\textwidth]{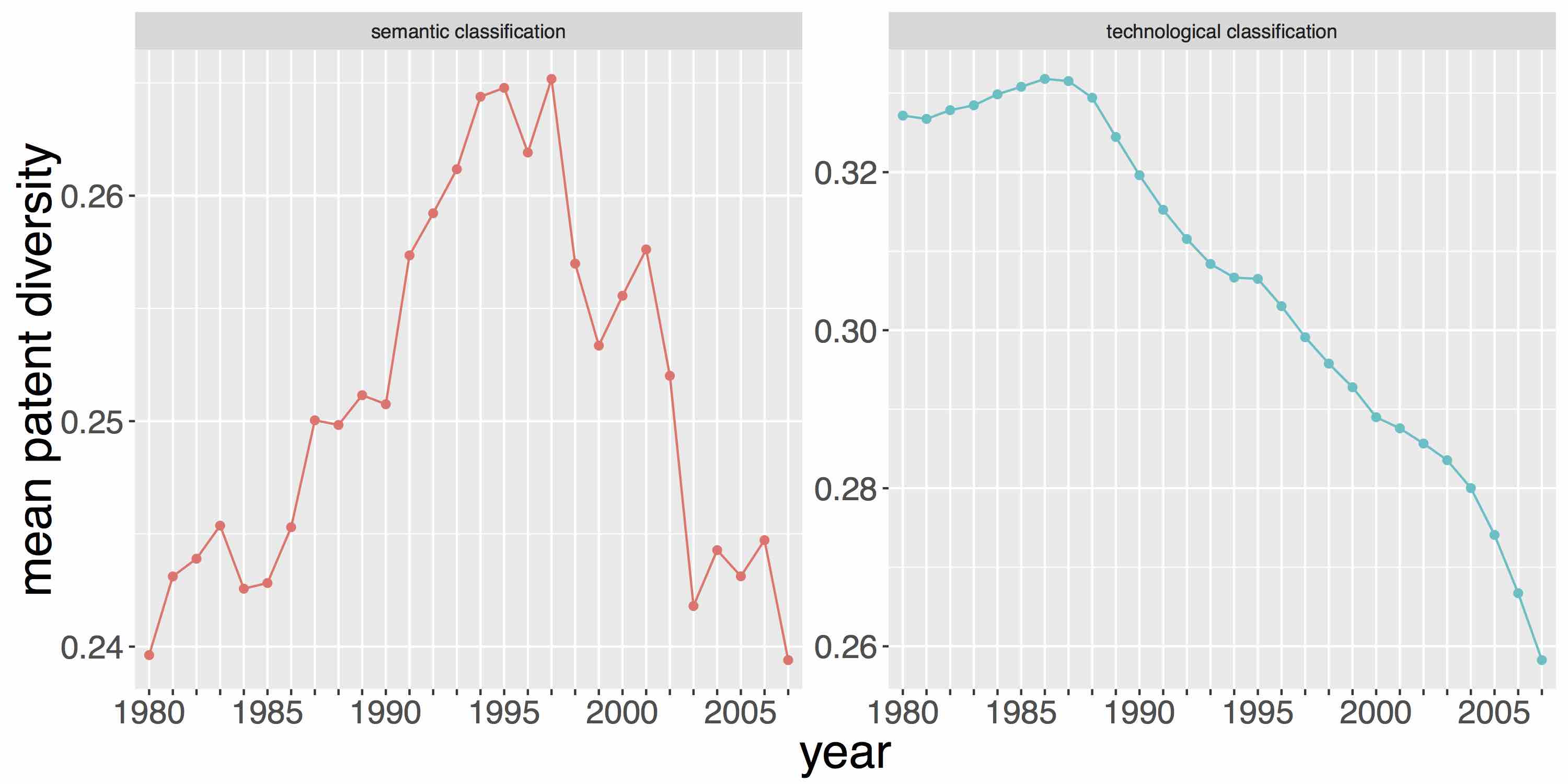}\\
\includegraphics[width=0.49\textwidth]{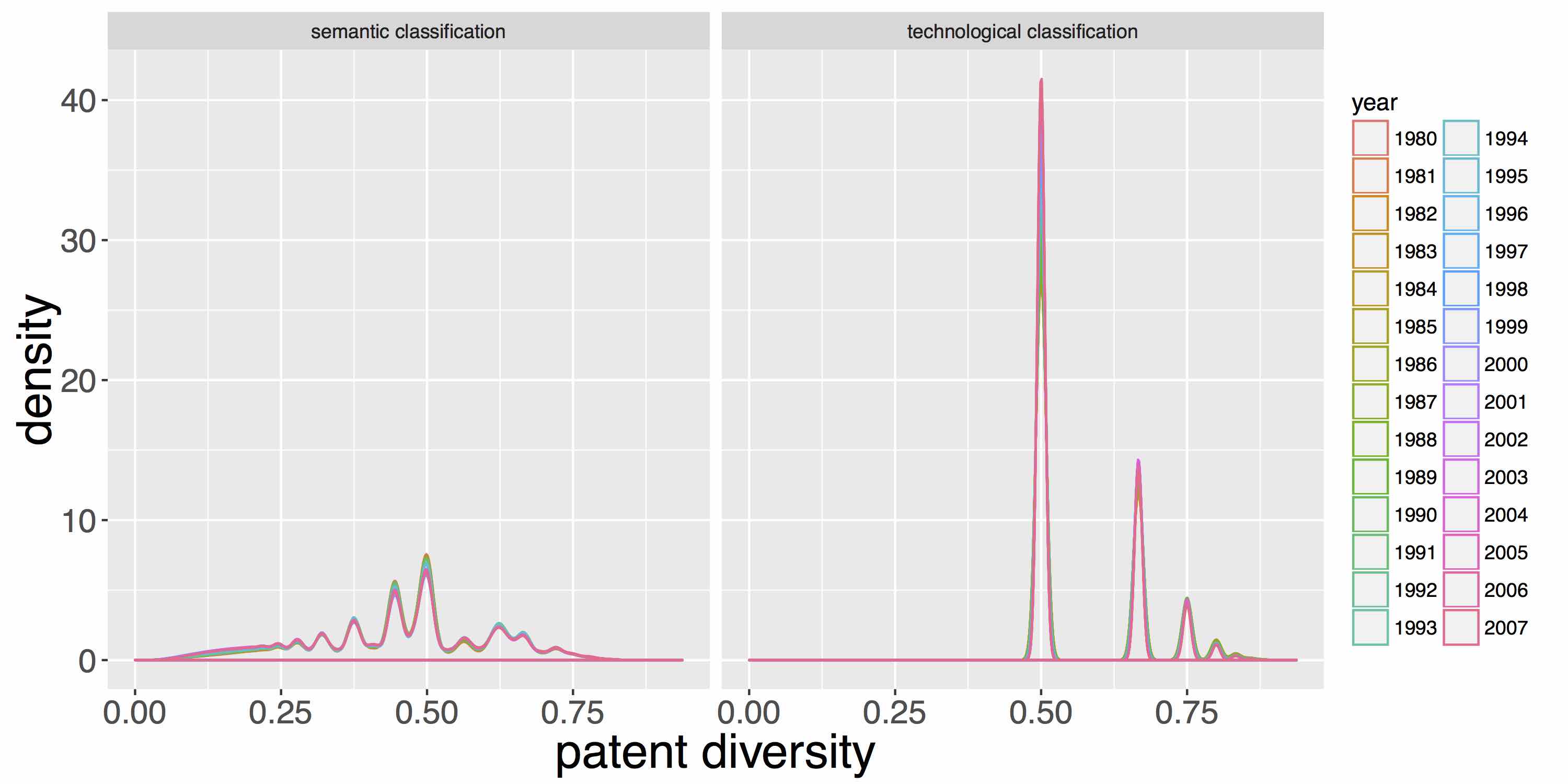}
\includegraphics[width=0.49\textwidth]{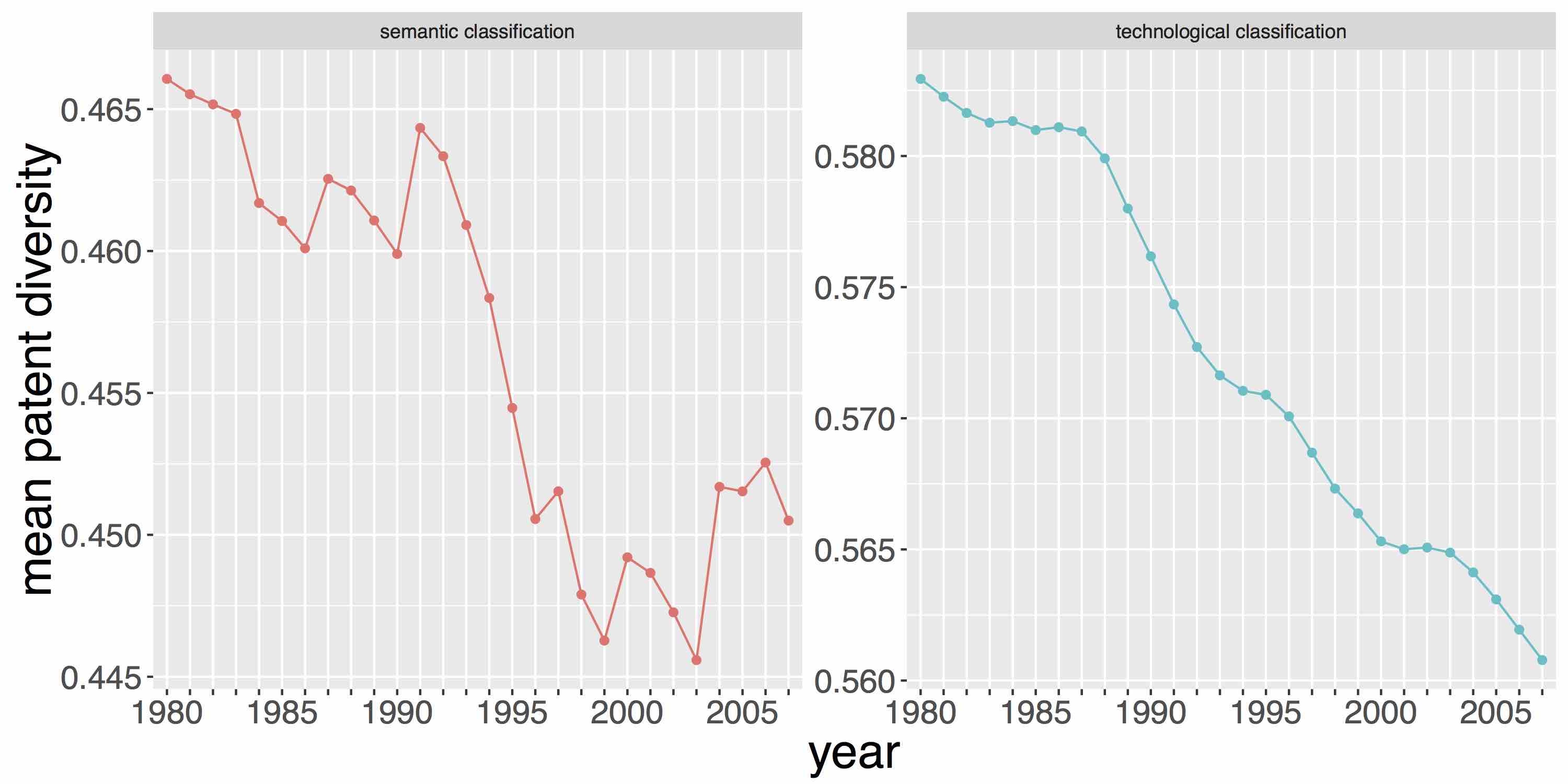}
\caption{
\textbf{Patent level diversities.} Distributions of diversities (Left column) and corresponding mean time-series (Right column) for $t=1980$ to $t=2007$ (with the corresponding time window $[t-4,t]$). The first row includes all classified patents, whereas the second row includes only patents with more than one class (i.e. patents with diversity greater than 0).
}
\label{fig:patent-level-orig}
\end{figure}

We show in Fig.~\ref{fig:patent-level-orig} the distribution over time of semantic and technological diversity with the corresponding mean time-series. This is carried with two different settings, namely including/not including patents with zero diversity (i.e. single class patents). We call other patents ``complicated patents'' in the following. First of all, the presence of mass in small probabilities for semantic but not technological diversity confirms that the semantic classification contains patent spread over a larger number of classes. More interestingly, a general decrease of diversity for complicated patents, both for semantic and technological classification systems, can be interpreted as an increase in invention specialization. This is a well-known stylized fact as documented in~\cite{ARCHIBUGI199279}. Furthermore, a qualitative regime shift on semantic classification occurs around 1996. This can be seen whether or not we include patents with zero diversity. The diversity of complicated patents stabilizes after a constant decrease, and the overall diversity begins to strongly decrease. This means that on the one hand the number of single class patents begins to increase and on the other hand complicated patents do not change in diversity. It can be interpreted as a change in the regime of specialization, the new regime being caused by more single-class patents.

\begin{figure}
\includegraphics[width=0.49\textwidth]{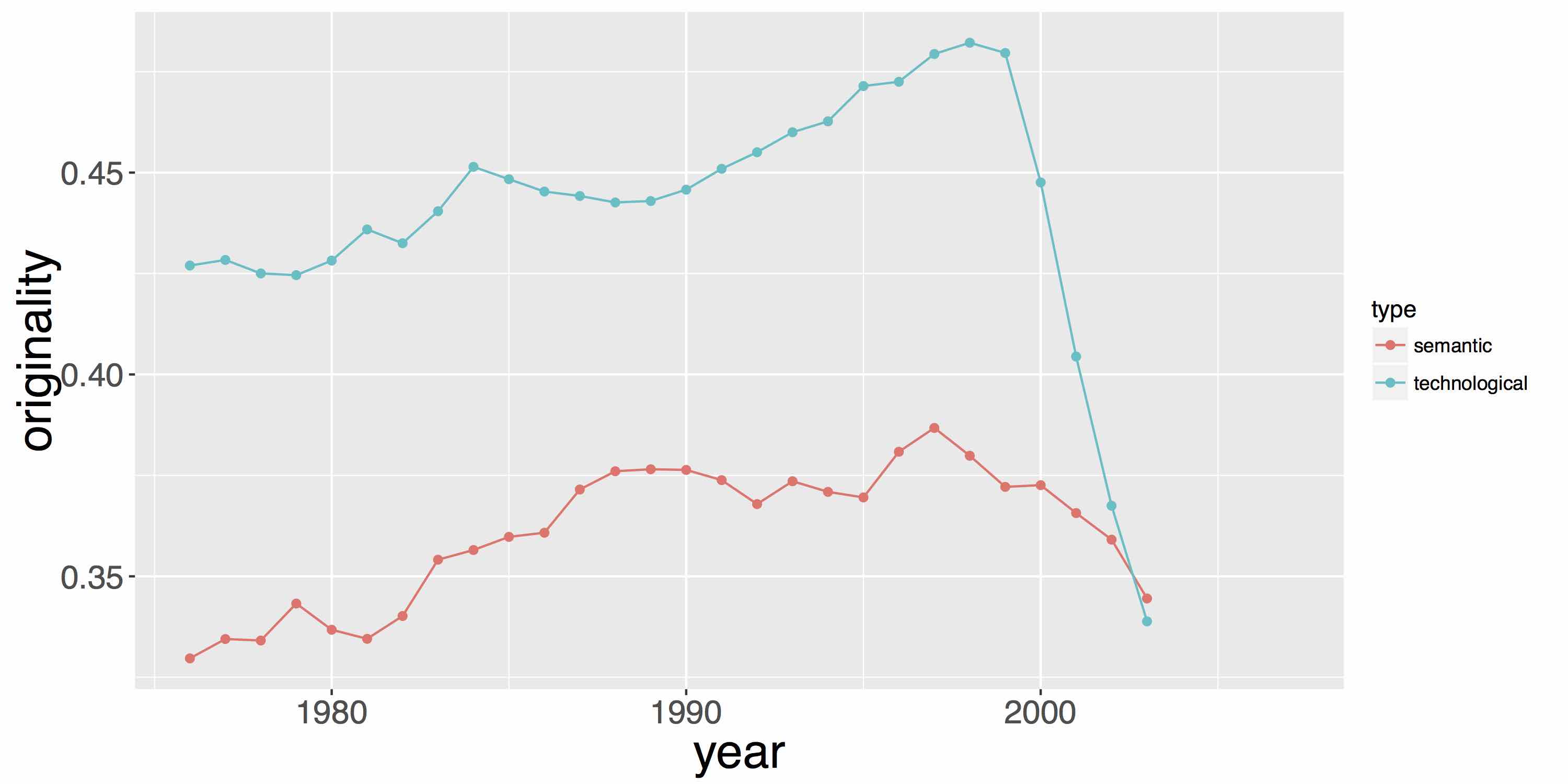}
\includegraphics[width=0.49\textwidth]{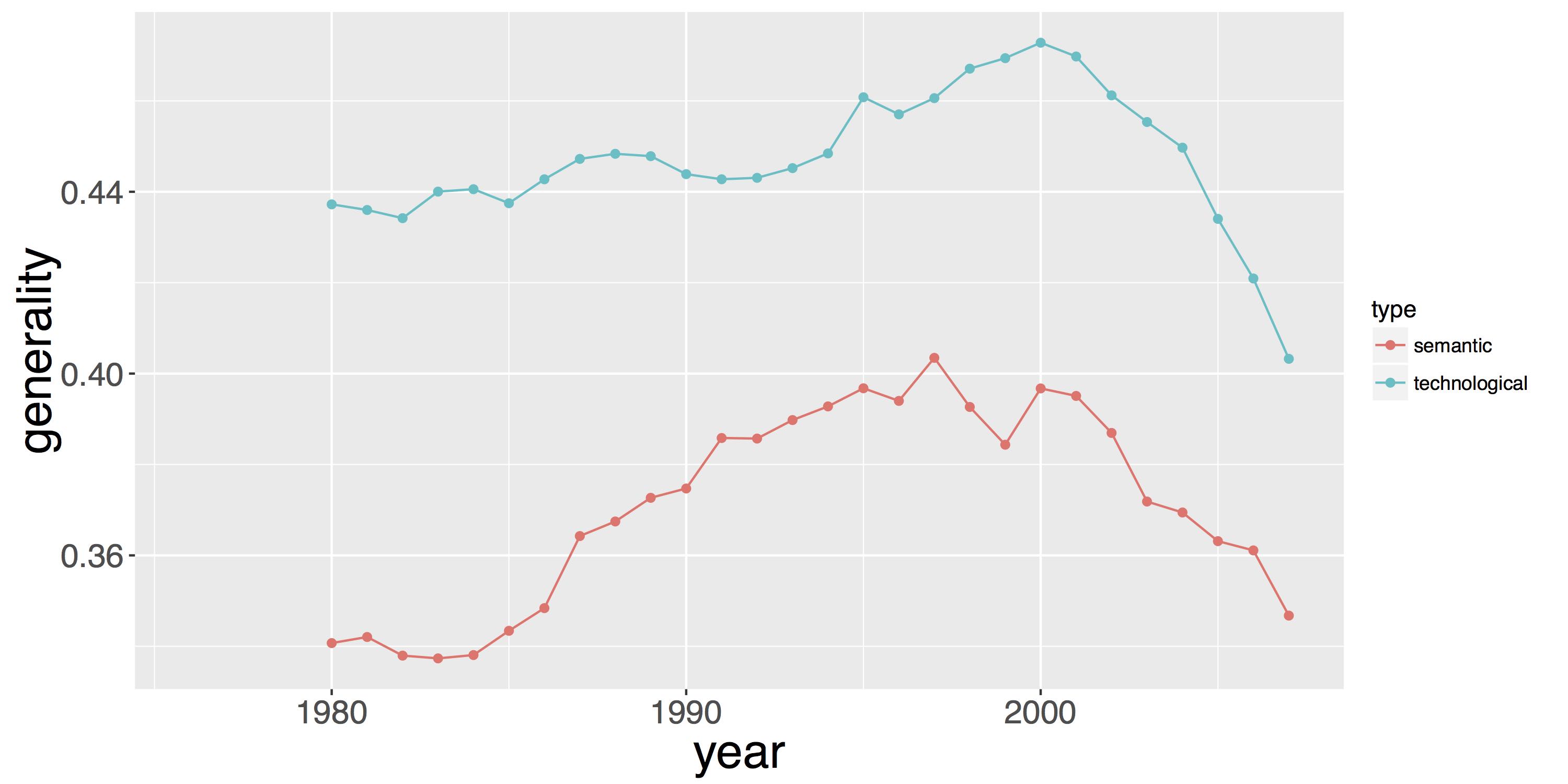}\\
\caption{\textbf{Patent level originality} (left hand side) and \textbf{generality} (right hand side)  for $t=1980$ to $t=2007$ (with the corresponding time window $[t-4,t]$) as defined in subsection \ref{subsec:orig-gene}. }
\label{fig:orig-gene}
\end{figure}

More commonly used in the literature are the measures of originality and generality. These measures follow the same idea than the above-defined diversity in quantifying the diversity of classes (whether technological or semantic) associated with a patent. But instead of looking at the patent's classes, they consider the classes of the patents that are cited or citing. Formally, the originality $O_i$ and the generality $G_i$ of a patent $i$ are defined as
\[
O_i^{(z)} = \displaystyle 1 - \sum_{j =1}^{N^{(z)}}{\left(\frac{\displaystyle \sum_{i' \in I_i}{p_{i'j}}}{\displaystyle \sum_{k =1}^{N^{(z)}}{\displaystyle \sum_{i' \in I_i}{p_{i'k}}}}\right)^2} \text{ and } G_i^{(z)} = \displaystyle 1 - \sum_{j =1}^{N^{(z)}}{\left(\frac{\displaystyle \sum_{i' \in \tilde{I}_i}{p_{i'j}}}{\displaystyle \sum_{k =1}^{N^{(z)}}{\displaystyle \sum_{i' \in \tilde{I}_i}{p_{i'k}}}}\right)^2}, 
\]
where $z \in \{tec, sem\}$, $I_i$ denotes the set of patents that are cited by the $i$th patent within a five year window (i.e. if the $i$th patent appears at year $t$, then we consider patents on $[t-T_0, t]$) when considering the originality and $\tilde{I}_i$ the set of patents that cite patent $i$ after less than five years (i.e. we consider patents on $[t ,t + T_0]$) in the case of generality. Note that the measure of generality is forward looking in the sense that $G_i^{(z)}$ used information that will only be available 5 years after patent applications. Both measures are lower on average based on semantic classification than on technological classification. Fig. \ref{fig:orig-gene} plots the mean value of $O_i^{(sem)}$, $O_i^{(tec)}$, $G_i^{(sem)}$ and $G_i^{(tec)}$.


\subsection{Classes overlaps} \label{subsec:overlaps}

A proximity measure between two classes can be defined by their overlap in terms of patents. 
Such measures could for example be used to construct a metrics between semantic classes. Intuitively, highly overlapping classes are very close in terms of technological content and one can use them to measure distance between two firms in terms of technology as done in~\cite{Bloom2005distance}. Formally, recalling the definition of $\left(p_{ij}\right)$ as the probability for the $i$th patent to belong to the $j$th class and $N_P$ as the number of patents it writes 
\begin{eqnarray}
\label{overlap}
Overlap_{jk} = \frac{1}{N_P}\cdot \sum_{i=1}^{N_P} p_{ij} p_{ik}. 
\end{eqnarray}

The overlap is normalized by patent count to account for the effect of corpus size: by convention, we assume the overlap to be maximal when there is only one class in the corpus. A corresponding relative overlap is computed as a set similarity measure in the number of patents common to two classes A and B, given by $o(A,B)=2\cdot \frac{\left|A\cap B\right|}{\left|A\right| + \left|B\right|}$.

\paragraph*{Intra-classification overlaps}

\begin{figure}[!ht]
\centering
\includegraphics[width=0.49\textwidth]{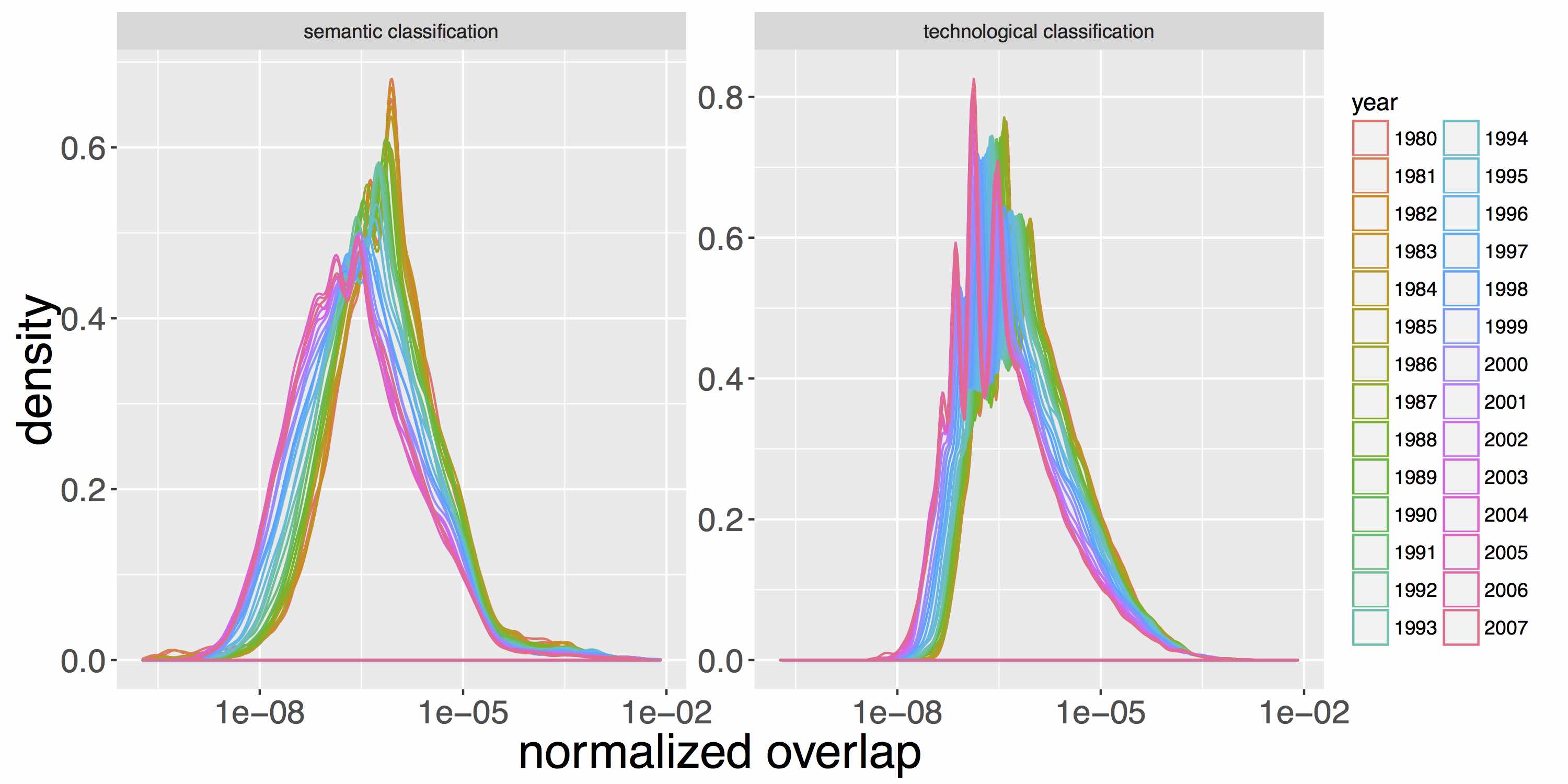}
\includegraphics[width=0.49\textwidth]{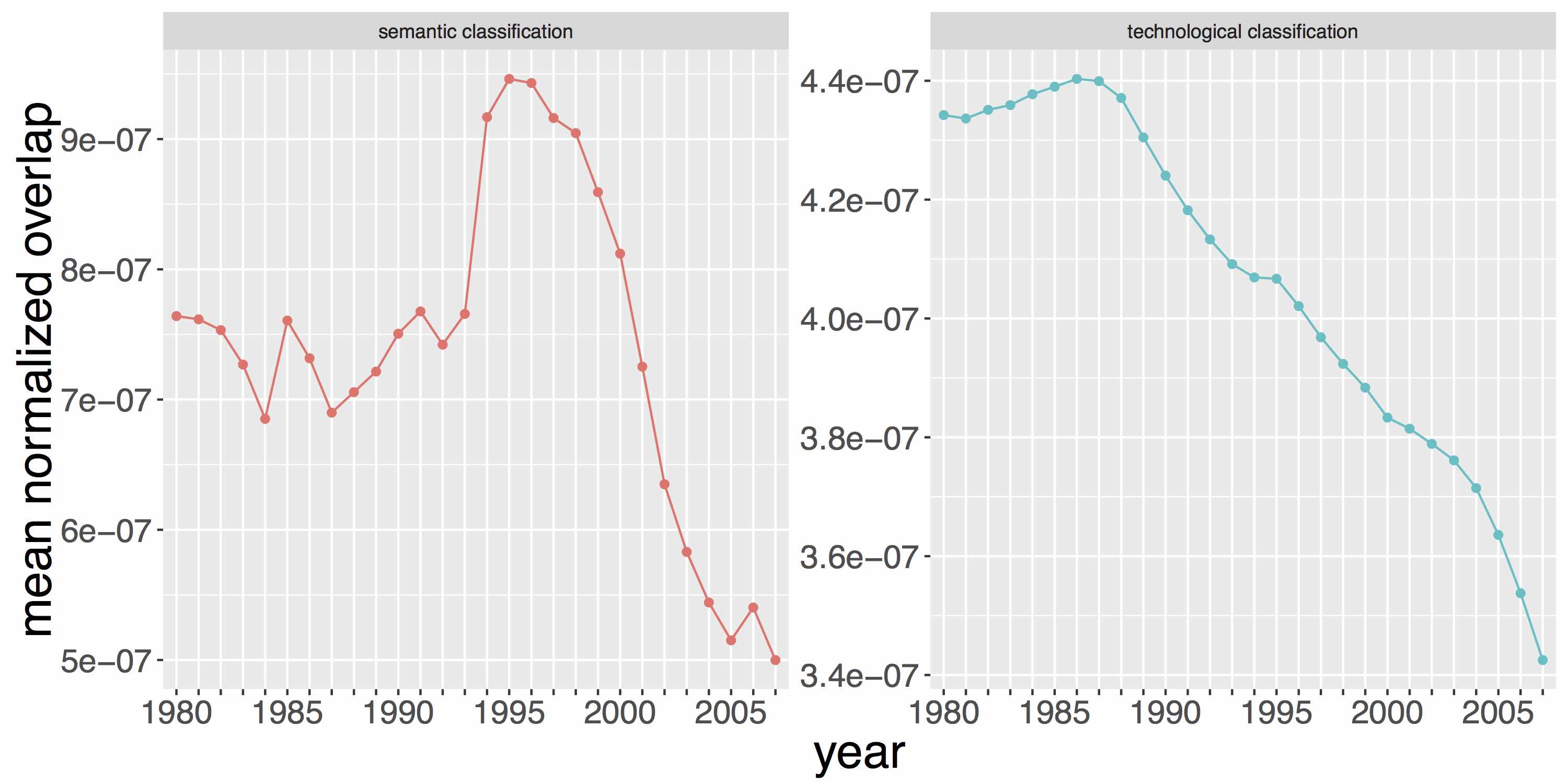} \\
\includegraphics[width=0.49\textwidth]{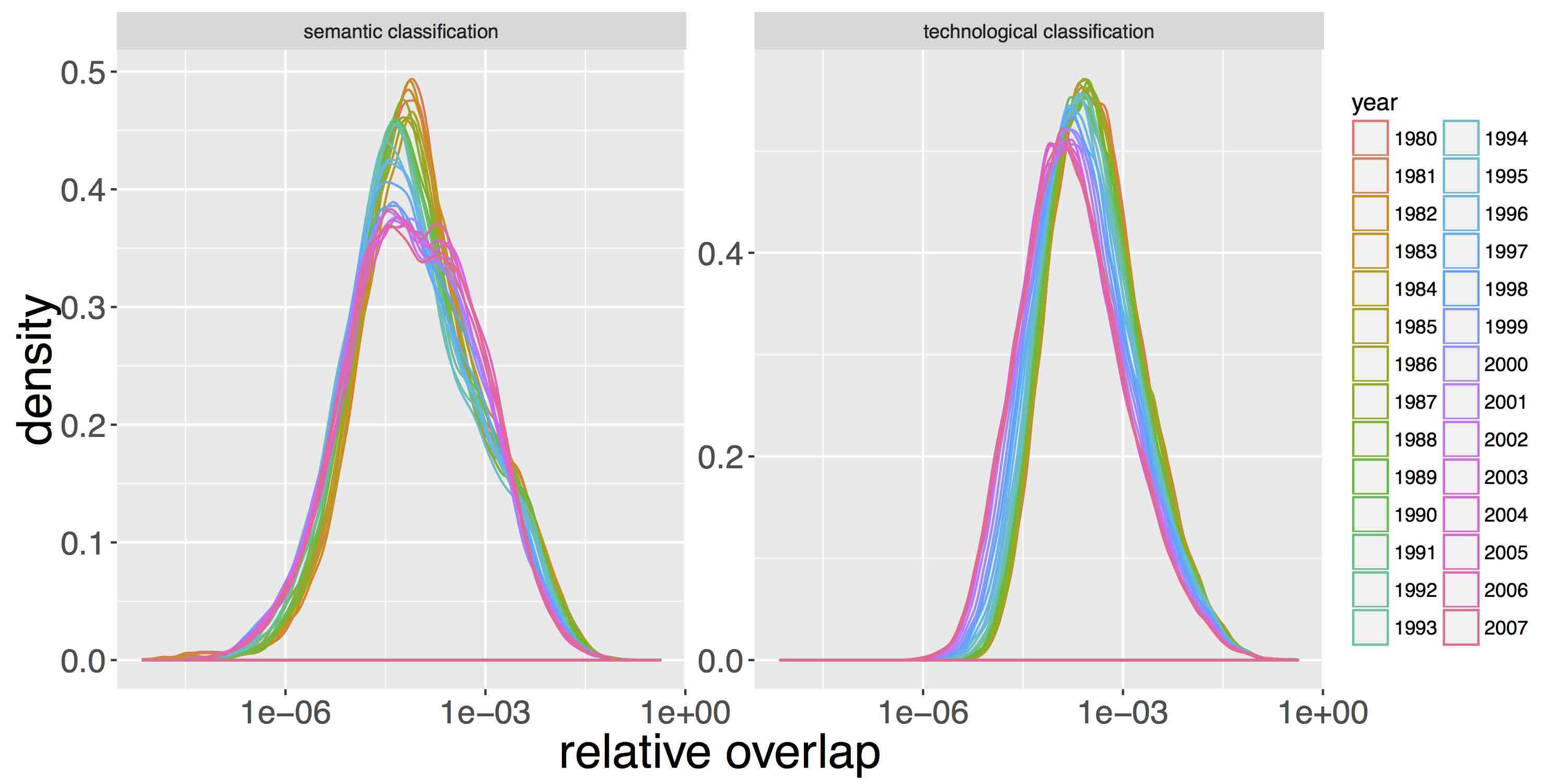}
\includegraphics[width=0.49\textwidth]{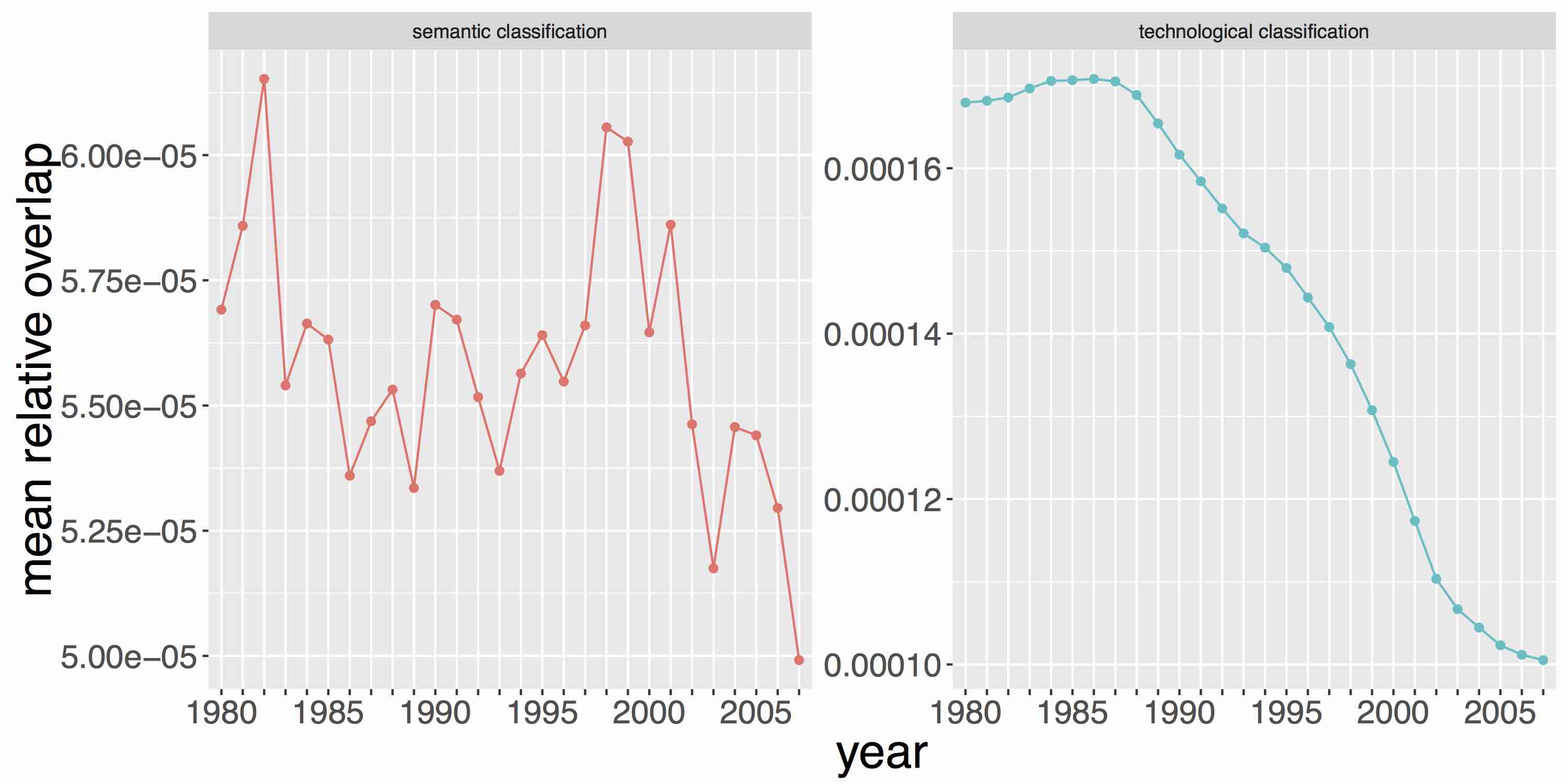}
\caption{
\textbf{Intra-classification overlaps.}
\textit{(Left column)} Distribution of overlaps $O_{ij}$ for all $i\neq j$ (zero values are removed because of the log-scale). \textit{(Right column)} Corresponding mean time-series. \textit{(First row)} Normalized overlaps. \textit{(Second row)} Relative overlaps.}
\label{fig:intra-classif-overlap}
\end{figure}

The study of distributions of overlaps inside each classification, i.e. between technological classes and between semantic classes separately, reveals the structural difference between the two classification methods, suggesting their complementary nature. Their evolution in time can furthermore give insights into trends of specialization. We show in Fig.~\ref{fig:intra-classif-overlap} distributions and mean time-series of overlaps for the two classifications. The technological classification globally always follow a decreasing trend, corresponding to more and more isolated classes, i.e. specialized inventions, confirming the stylized fact obtained in previous subsection. For semantic classes, the dynamic is somehow more intriguing and supports the story of a qualitative regime shift suggested before. Although globally decreasing as technological overlap, normalized (resp. relative) mean overlap exhibits a peak (clearer for normalized overlap) culminating in 1996 (resp. 1999). Looking at normalized overlaps, classification structure was somewhat stable until 1990, then strongly increased to peak in 1996 and then decrease at a similar pace up to now. Technologies began to share more and more until a breakpoint when increasing isolation became the rule again. An evolutionary perspective on technological innovation~\cite{ziman2003technological} could shed light on possible interpretations of this regime shift: as species evolve, the fitness landscape first would have been locally favorable to cross-insemination, until each fitness reaches a threshold above which auto-specialization becomes the optimal path. It is very comparable to the establishment of an ecological niche~\cite{holland2012signals}, the strong interdependency originating here during the mutual insemination resulting in a highly path-dependent final situation. 

\paragraph*{Inter-classification overlaps}

\begin{figure}[!ht]
\includegraphics[width=0.49\textwidth]{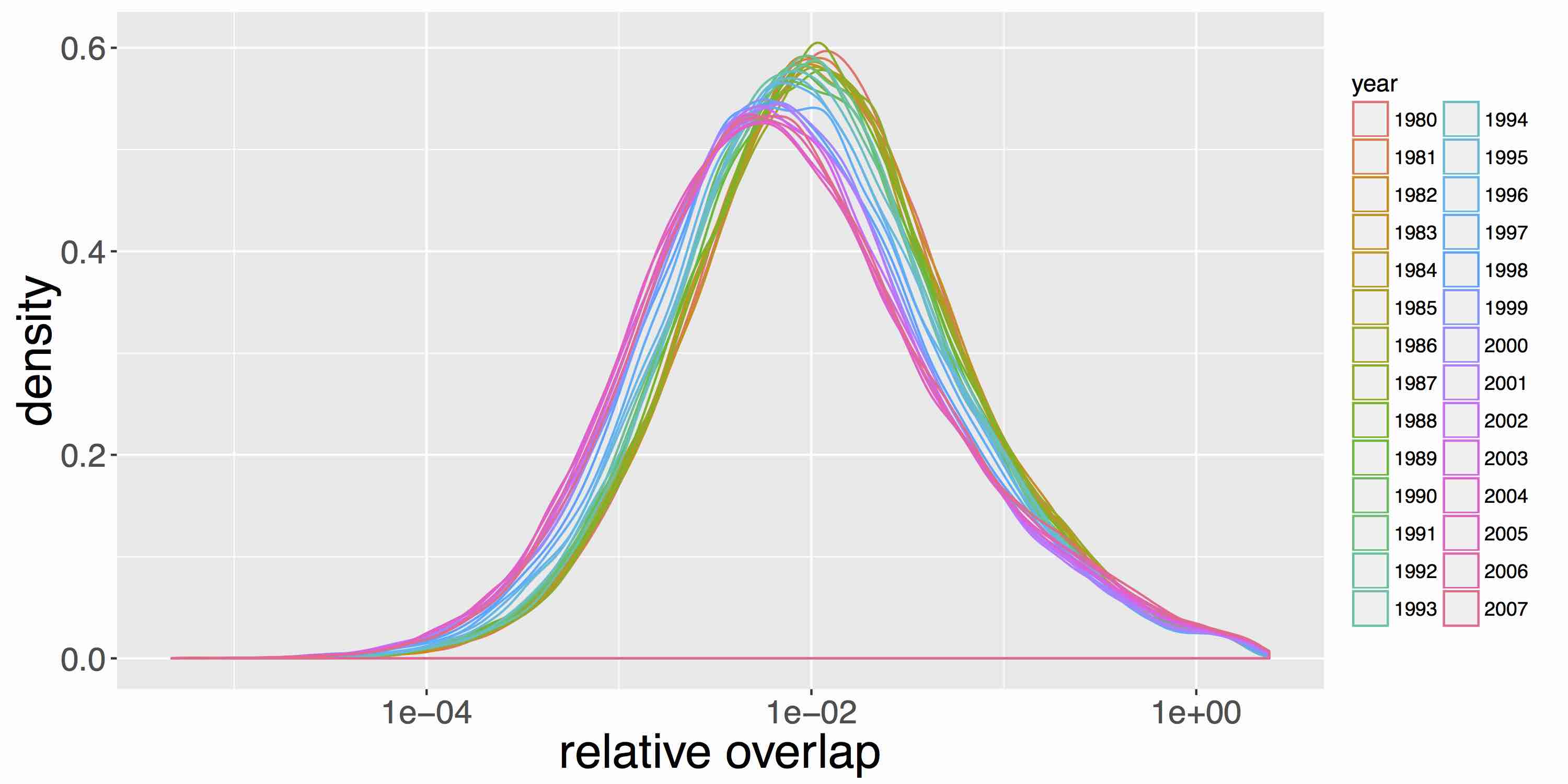}
\includegraphics[width=0.49\textwidth]{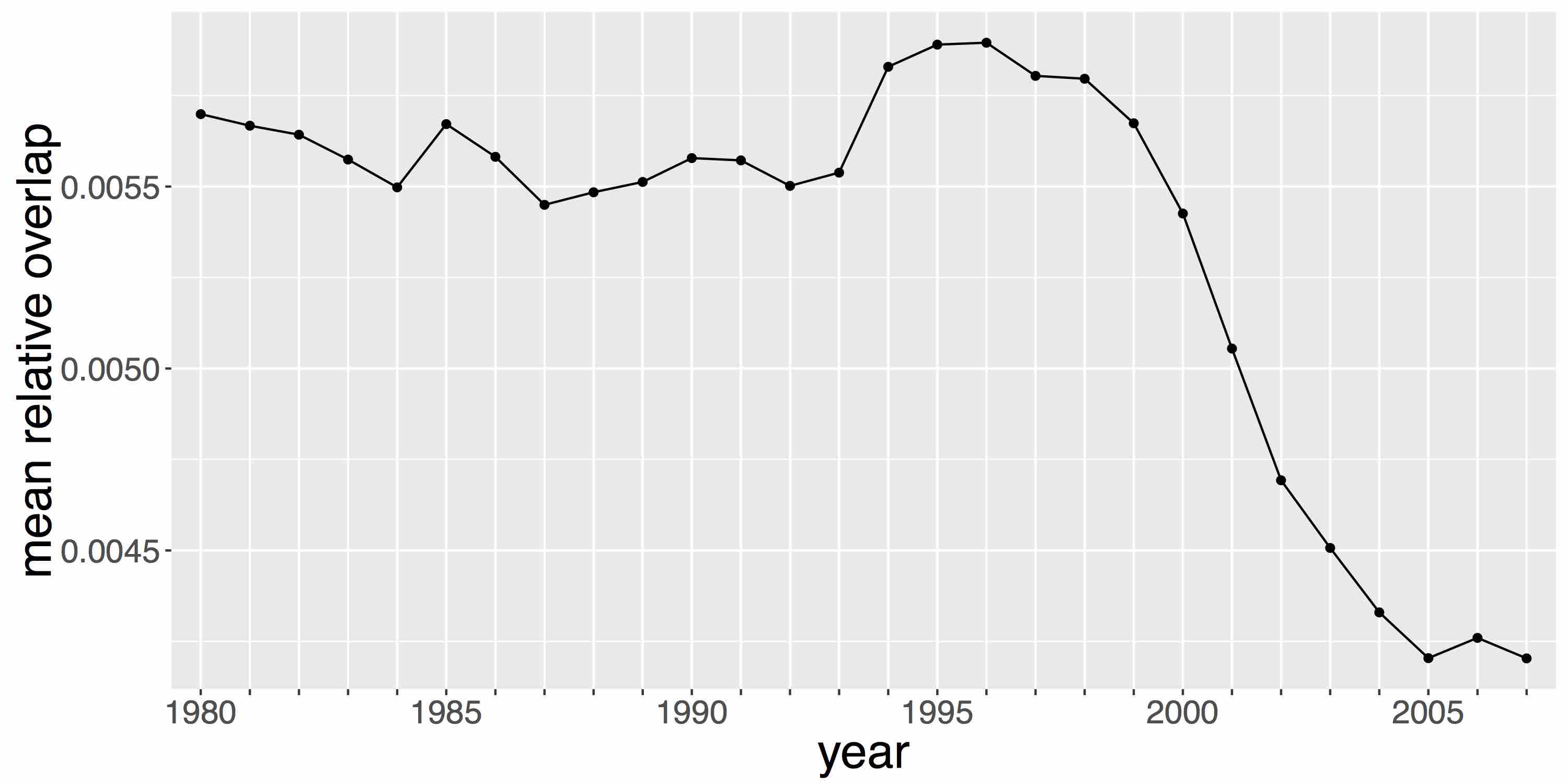}
\caption{\textbf{Distribution of relative overlaps between classifications.} (Left) Distribution of overlaps at all time steps; (Right) Corresponding mean time-series. The decreasing trend starting around 1996 confirms a qualitative regime shift in that period.}
\label{fig:inter-classif-overlap}
\end{figure}

Overlaps \emph{between} classifications are defined as in (\ref{overlap}), but with $j$ standing for the $j$th technological class and $k$ for the $k$th semantic class: $p_{ij}$ are technological probabilities and $p_{ik}$ semantic probabilities. They describe the relative correspondence between the two classifications and are a good indicator to spot relative changes, as shown in Fig.~\ref{fig:inter-classif-overlap}. Mean inter-classification overlap clearly exhibits two linear trends, the first one being constant from 1980 to 1996, followed by a constant decrease. Although difficult to interpret directly, this stylized fact clearly unveils a change in the \emph{nature} of inventions, or at least in the relation between content of inventions and technological classification. As the tipping point is at the same time as the ones observed in the previous section and since the two statistics are different, it is unlikely that this is a mere coincidence. Thus, these observations could be markers of a hidden underlying structural changes in processes.

\subsection{Citation Modularity}
\label{citationmodularity}

An exogenous source of information on relevance of classifications is the citation network described in Section \ref{sub:citation}. The correspondence between citation links and classes should provide a measure of accuracy of classifications, in the sense of an external validation since it is well-known that citation homophily  is expected to be quite high (see, e.g, ~\cite{AAKnetwork2016}). This section studies empirically modularities of the citation network regarding the different classifications. To corroborate the obtained results, we propose to look at a more rigorous framework in Section \ref{statisticalmodel}. Modularity is a simple measure of how communities in a network are well clustered (see \cite{clauset2004finding} for the accurate definition). Although initially designed for single-class classifications, this measure can be extended to the case where nodes can belong to several classes at the same time, in our case with different probabilities as introduced in \cite{nicosia2009extending}. The simple directed modularity is given in our case by
\[
Q_d^{(z)} = \displaystyle \frac{1}{N_P}\sum_{1\leq i,j\leq N_P}\left[A_{ij} - \frac{k_{i}^{in}k_{j}^{out}}{N_P}\right]\delta(c_i,c_j),
\]
with $A_{ij}$ the citation adjacency matrix (i.e. $A_{ij} = 1$ if there is a citation from the $i$th patent to the $j$th patent, and $A_{ij}=0$ if not), $k_i^{in}=\left| I_i\right|$ (resp. $k_i^{out}= \left|\tilde{I}_i \right|$) in-degree (resp. out-degree) of patents (i.e. the number of citations made by the $i$th patent to others and the number of citations received by the $i$th patent). $Q_d$ can be defined for each of the two classification systems: $z \in \{tec, sem\}$. If $z=tec$, $c_i$ is defined as the main patent class, which is taken as the first class whereas if $z=sem$, $c_i$ is the class with the largest probability.

\begin{figure}[!ht]
\centering
\includegraphics[width=0.49\textwidth]{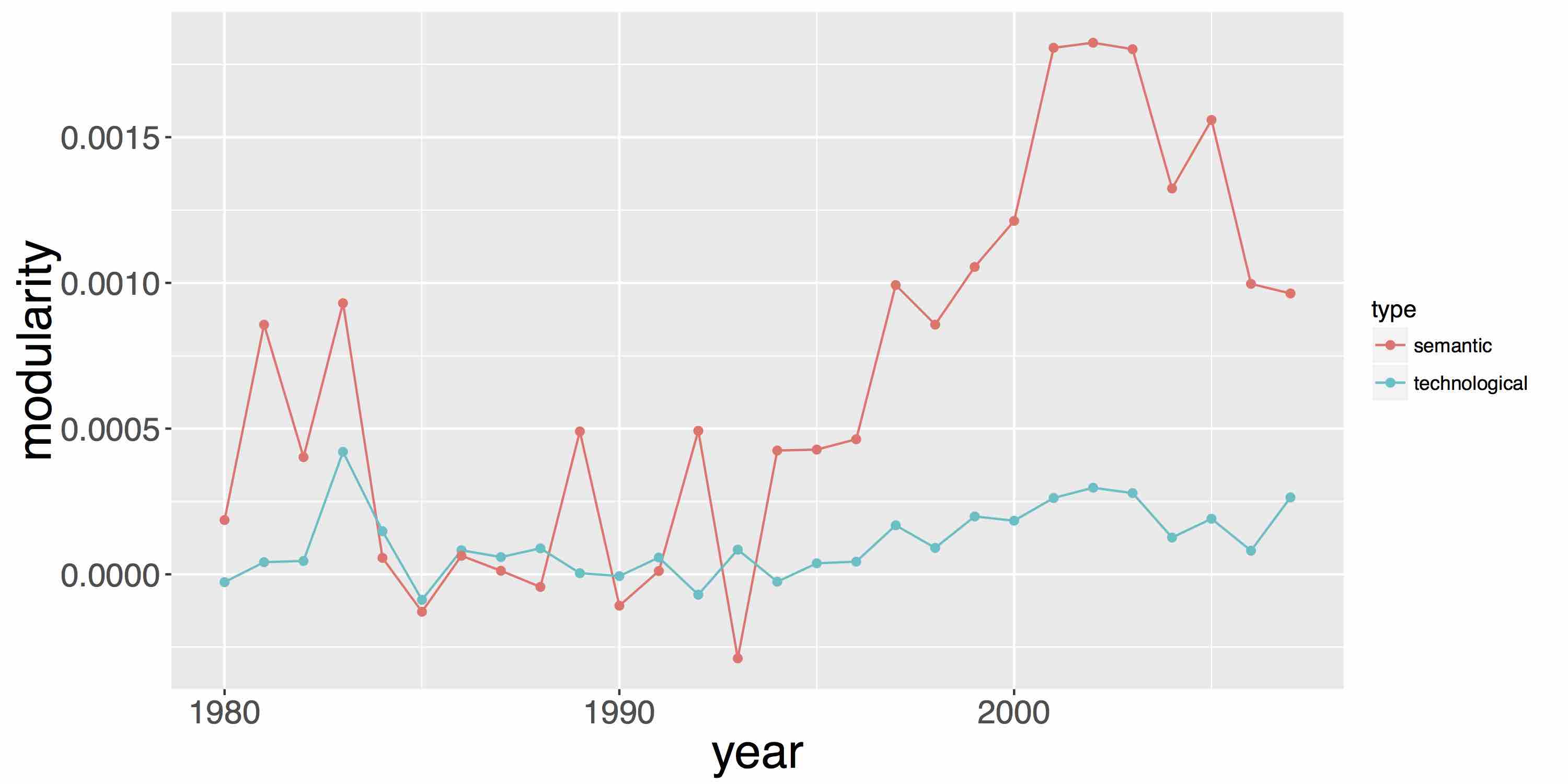}
\includegraphics[width=0.49\textwidth]{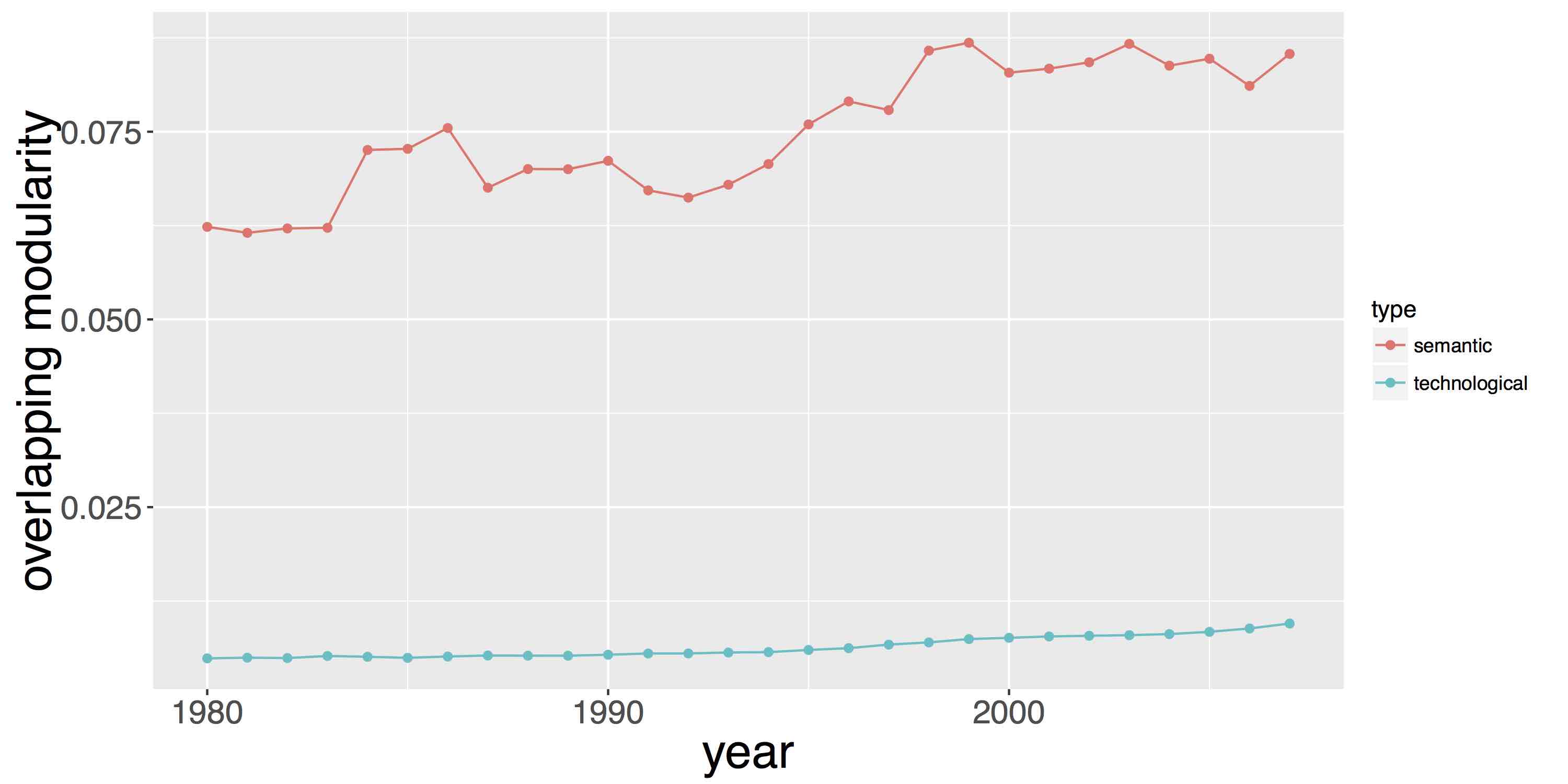}
\caption{\textbf{Temporal evolution of semantic and technological modularities of the citation network.} (Left) Simple directed modularity, computed with patent main classes (main technological class and semantic class with larger probability). (Right) Multi-class modularity, computed following~\cite{nicosia2009extending} }
\label{fig:modularities}
\end{figure}

Multi-class modularity in turns is given by

\[
\displaystyle Q_{ov}^{(z)} = \frac{1}{N_P} \sum_{c = 1}^{N^{(z)}} \sum_{1\leq i,j \leq N_P}\left[F(p_{ic},p_{jc})A_{ij} - \frac{\beta_{i,c}^{out}k_i^{out}\beta_{j,c}^{in}k_j^{in}}{N_P}\right],
\]
where
\[
 \beta_{i,c}^{out} =   \frac{1}{N_P} \displaystyle \sum_j F(p_{ic},p_{jc}) \text{ and } \beta_{j,c}^{in} =  \frac{1}{N_P} \displaystyle \sum_i F(p_{ic},p_{jc}).
\]
We take $F(p_{ic},p_{jc}) = p_{ic}\cdot p_{jc}$ as suggested in \cite{nicosia2009extending}. Modularity is an aggregated measure of how the network deviates from a null model where links would be randomly made according to node degree. In other words it captures the propensity for links to be inside the classes. Overlapping modularity naturally extends simple modularity by taking into account the fact that nodes can belong simultaneously to many classes.
We document in Fig.~\ref{fig:modularities} both simple and multi-class modularities over time. For simple modularity, $Q_d^{(tec)}$ is low and stable across the years whereas $Q_d^{(sem)}$ is slightly greater and increasing. These values are however low and suggest that single classes are not sufficient to capture citation homophily. Multi-class modularities tell a different story. First of all, both classification modularities have a clear increasing trend, meaning that they become more and more adequate with citation network. The specializations revealed by both patent level diversities and classes overlap is a candidate explanation for this growing modularities. Secondly, semantic modularity dominates technological modularity by an order of magnitude (e.g. 0.0094 for technological against 0.0853 for semantic in 2007) at each time. This discrepancy has a strong qualitative significance. Our semantic classification fits better the citation network when using multiple classes. As technologies can be seen as a combination of different components as shown by~\cite{Youn:2015fk}, this heterogeneous nature is most likely better taken into account by our multi-class semantic classification.

\subsection{Statistical Model}
\label{statisticalmodel}
In this section, we develop a statistical model aimed at quantifying performance of both technological and semantic classification systems. In particular, we aim at corroborating findings obtained in Section \ref{citationmodularity}. The mere difference between this approach and the citation modularity approach lies in the choice of the underlying model, and the according quantities of interest. In addition for the semantic approach, we want to see if when restricting to patents with higher probabilities to belong to a class, we obtain better results. To do that, we choose to look at within class citations proportion (for both technological and semantic approaches). We provide two obvious reasons why we choose this. First, the citations are commonly used as a proxy for performance as mentioned in Section \ref{citationmodularity}. Second, this choice is ``statistically fair'' in the sense that both approaches have focused on various goals and not on maximizing directly the within class proportion. 
Nonetheless, the within class proportion is too sensitive to the distribution of the shape of classes. For example, a dataset where patents for each class account for 10\% of the total number of patents will mechanically have a better within class proportion than if each class accounts for only 1\%. Consequently, an adequate statistical model, which treats datasets fairly regardless of their distribution in classes, is needed. This effort ressembles to the previous study of citation modularity, but is complementary since the model presented here can be understood as an elementary model of citation network growth. Furthermore, the parameters fitted here can have a direct interpretation as a citation probability.

We need to introduce and recall some notations. We consider a specific window of observations $\big[t- T_0, t \big]$, and we define $Z$ the number of patents which appeared during that time window. We let $t_1, \cdots, t_Z$ their corresponding appearance date by chronological order, which for simplicity are assumed to be such that $t_1 < \cdots < t_Z$. For each patent $i=1, \cdots, Z$ we consider $C_i$ the number of distinctive couples \{cited patent, cited patent's class\} made by the $i$th patent (for instance if the $i$th patent has only made one citation and that the cited patent is associated with three classes, then $C_i = 3$). Let $z \in \{tec, sem\}$, we define $N_{i}^{(z)}$  the number of patents associated to at least one of the $i$th classes at time $t_{i-1}$. For $l = 1, \cdots, C_i$ we consider the variables $B_{l,i}$, which equal $1$ if the cited patent's class is also common to the $i$th patent. We assume that $B_{l,i}$ are independent of each other and conditioned on the past follow Bernoulli variables 
$$B \Big( \min \Big\{ 1, \frac{N_{i}^{(z)}}{i-1} + \theta^{(z)} \Big\} \Big),$$ 
where the parameter $0 \leq \theta^{(z)} \leq 1$ indicates the propensity for any patent to cite patents of its own technological or semantic class. When $\theta^{(z)} = 0$, the probability of citing patents from its own class is simply $N_{i}^{(z)}(i-1)^{-1}$, which corresponds to the observed proportion of patents which belong to at least one of the $i$th patent's classes. Thus this corresponds to the estimated probability of citing one patent if we assume that the probability of citing any patent $k=1, \cdots, i-1$ is uniformly distributed, which could be a reasonable assumption if classes were assigned randomly and independently from patent abstract contents. Conversely if $\theta^{(z)} = 1$, we are in the case of a model where there are 100\% of within class citations. A reasonable choice of $\theta^{(z)}$ lies between those two extreme values. Finally, we assume that the number of distinctive couples $C_i$ are a sequence of independent and identically distributed random variables following the discrete distribution $C$, and also independent from the other quantities.

We estimate $\theta^{(z)}$ via maximum likelihood, and obtain the corresponding maximum likelihood estimator (MLE) $\hat{\theta}^{(z)}$. The likelihood function, along with the standard deviation expression and details about the test, can be found in~\nameref{sectionSI}. The fitted values, standard errors and p-values corresponding to the statistical test $\theta^{(sem)} = \theta^{(tec)}$ (with corresponding alternative hypothesis $\theta^{(sem)} > \theta^{(tec)}$) on non-overlapping blocks from the period 1980-2007\footnote{Note that the estimation included patents up until 2010 in the period 2006-2007 and not the patents from 1980 in the period 1980-1985 for homogeneity in size with other periods. This doesn't affect the significativity of the results.} are reported on Table \ref{summary}. Semantic values are reported for four different chosen thresholds $p^{-}=.04, .06, .08, .1$. It means that we restricted to the couples ($i$th patent, $j$th class) such that $p_{ij} \geq p^{-}$. 

The choice of considering non-overlapping blocks (instead of overlapping blocks) is merely statistical. Ultimately, our interest is in the significance of the test over the whole period 1980-2007. Thus, we want to compute a global p-value. This can be done considering the local p-values (by local, we mean for instance computed on the period 2001-2005) assuming independence between them. This assumption is reasonable only if the blocks are non-overlapping. All of this can be found in~\nameref{sectionSI}. Finally, note that from a statistical perspective, including overlapping blocks wouldn't yield more information.

The values reported in Table \ref{summary} are overwhelmingly against the null hypothesis. The global estimates of $\theta^{(sem)}$ are significantly bigger than the estimate of $\theta^{(tec)}$ for all the considered thresholds. Although the corresponding p-values (which are also very close to 0) are not reported, it is also quite clear that the bigger the threshold, the higher the corresponding $\theta^{(sem)}$ is estimated. This is consistently seen for any period, and significant for the global period. This seems to indicate that when restricting to the couples (patent, class) with high semantic probability, the propension to cite patents from its own class $\theta^{(sem)}$ is increasing. We believe that this might provide extra information to patent officers when making their choice of citations. Indeed, they could look first to patents which belong to the same semantic class, especially when patents have high probability semantic values.  

Note that the introduced model can be seen as a simple model of citations network growth conditional to a classification, which can be expressed as a stochastic block model (e.g. \cite{decelle2011asymptotic}, \cite{valles2016multilayer}). The parameters are estimated computing the corresponding MLE. In view of~\cite{2016arXiv160602319N}, this can be thought as equivalent to maximizing modularity measures.

\begin{table}[!ht]
\centering
\caption{\label{summary} Estimated values of $\theta^{(tec)}$ and $\theta^{(sem)}$ and corresponding standard errors obtained from a Maximum Likelihood estimator as presented in section \ref{statisticalmodel}.}
\label{stdQMLE}
\begin{tabular}{@{}rccccccccc@{}}
\toprule
\toprule
\multicolumn{1}{l}{Approach} & \multicolumn{1}{l}{Estimated Value} & \multicolumn{1}{l}{st. er.} & \multicolumn{1}{l}{p-value} \\ \toprule
\multicolumn{4}{l}{1980-1985 period} \\
technological & .664 &.008&\\
semantic $p^{-} = .04$ & .741 &.047 & .053\\
semantic $p^{-} = .06$ & .799 &.081 & .049\\
semantic $p^{-} = .08$ & .828 &.126 & .097\\
semantic $p^{-} = .10$ & .834 &.166 & .153\\
\multicolumn{4}{l}{1986-1990 period} \\
technological & .634 &.007&\\
semantic $p^{-} = .04$ & .703 &.022 & .001\\
semantic $p^{-} = .06$ & .768 &.040 & .0004\\
semantic $p^{-} = .08$ & .804 &.069 & .007\\
semantic $p^{-} = .10$ & .832 &.114 & .041\\
\multicolumn{4}{l}{1991-1995 period} \\
technological & .619 &.006&\\
semantic $p^{-} = .04$ & .655 &.009 & .0004\\
semantic $p^{-} = .06$ & .713 &.017 & 9e-08\\
semantic $p^{-} = .08$ & .731 &.025 & 7e-06\\
semantic $p^{-} = .10$ & .750 &.037 & 9e-06\\
\multicolumn{4}{l}{1996-2000 period} \\
technological & .551 &.003&\\
semantic $p^{-} = .04$ & .585 &.002 & $\approx 0$\\
semantic $p^{-} = .06$ & .638 &.004 & $\approx 0$\\
semantic $p^{-} = .08$ & .660 &.006 & $\approx 0$\\
semantic $p^{-} = .10$ & .686 &.008 & $\approx 0$\\
\multicolumn{4}{l}{2001-2005 period} \\
technological & .567 &.003&\\
semantic $p^{-} = .04$ & .621 &.004 & $\approx 0$\\
semantic $p^{-} = .06$ & .676 &.007 & $\approx 0$\\
semantic $p^{-} = .08$ & .701 &.010 & $\approx 0$\\
semantic $p^{-} = .10$ & .710 &.013 & $\approx 0$\\
\multicolumn{4}{l}{2006-2007 period} \\
technological & .600 &.007&\\
semantic $p^{-} = .04$ & .683 &.016 & 1e-06\\
semantic $p^{-} = .06$ & .732 &.025 & 2e-07\\
semantic $p^{-} = .08$ & .760 &.036 & 6e-06\\
semantic $p^{-} = .10$ & .782 &.048 &9e-05\\
\multicolumn{4}{l}{1980-2007 global period} \\
technological & .606 &.002&\\
semantic $p^{-} = .04$ & .665 &.009 & 8e-11\\
semantic $p^{-} = .06$ & .721 &.017 & 9e-12\\
semantic $p^{-} = .08$ & .747 &.025 & 9e-09\\
semantic $p^{-} = .10$ & .782 &.035 & 3e-07\\
\bottomrule
\end{tabular}
\end{table}

\section{Conclusion \label{discussion}}

The main contribution of this study was twofold. First we have defined how we built a network of patents based on a classification that uses semantic information from abstracts. We have shown that this classification share some similarities with the traditional technological classification, but also have distinct features. Second, we provide researchers with materials resulting from our analysis, which includes: (i) a database linking each patent with its set of semantic classes and the associated probabilities; (ii) a list of these semantic classes with a description based on the most relevant keywords; (iii) a list of patent with their topological properties in the semantic network (centrality, frequency, degree etc...). The availability of these data suggests new avenues for further research.

A first potential application is to use the patents' topological measures inherited from their relevant keywords. The fact that these measures are backward-looking and immediately available after the publication of the patent information is an important asset. It would for example be very interesting to test their predicting power to assess the quality of an innovation, using the number of forward citations received by a patent, and subsequently the future effect on the firm's market value. 

Regarding firm innovative strategy, a second extension could be to study trajectories of firms in the two networks: technological and semantic. Merging these information with data on the market value of firms can give a lot of insight about the more efficient innovative strategies, about the importance of technology convergence or about acquisition of small innovative firms. It will also allow to observe innovation pattern over a firm life cycle and how this differ across technology field.

A third extension would be to use dig further into the history of innovation. USPTO patent data have been digitized from the first patent in July 1790. However, not all of them contain a text that is directly exploitable. We consider that the quality of patent's images is good enough to rely on Optical Character Recognition techniques to retrieve plain text from at least 1920. With such data, we would be able to extend our analysis further back in time and to study how technological progress occurs and combines in time. \cite{akcigit2013mechanics} conduct a similar work by looking at recombination and apparition of technological subclasses.
Using the fact that communities are constructed yearly, one can construct a measure of proximity between two successive classes. This could give clear view on how technologies converged over the year and when others became obsolete and replaced by new methods.


\section*{Supporting Information \label{sectionSI}}

\subsection*{S1 Text : Definition of utility patent}

\textbf{Describes with more details the definition of patents and  context.}

\subsection*{S2 Text : Data collection procedure}

\textbf{Detailed description of data collection}

\subsection*{S3 File : Semantic Network Visualization}

\textbf{Vector file of the semantic network (Fig.\ref{fig:rawnetwork})} available at\\
 \url{http://37.187.242.99/files/public/network.svg}

\subsection*{S4 Text : Network Sensitivity Analysis\label{app:sensitivity}}

\textbf{Extended figures for Network Sensitivity Analysis}

\subsection*{S5 Text : Statistical definitions and derivations}

\textbf{Extended definitions and derivations for the statistical model}



%
%
%


\newpage

\section*{S1 Text : Definition of utility patent} 

A utility patent at the USPTO is a document providing intellectual property and protection of an invention. It excludes others to making, using, or selling the invention the same invention in the United States in exchange for a disclosure of the patent content. The protection is granted for 20 years since 1995 (it was 17 years before that from 1860) starting from the year the patent application was filled, but can be interrupted before if its owner fails to pay the maintenance fees due after 3.5, 7.5 and 11.5 years. Utility patents are by far the most numerous, with more than 90\% of the total universe of USPTO patents.\footnote{Other categories are Plant patents, Design patents and Reissue patents.} According to the Title 35 of the United States Codes (35 USC) section 101: \textit{``Whoever invents or discovers any new and useful process, machine, manufacture, or composition of matter, or any new and useful improvement thereof, may obtain a patent therefor, subject to the conditions and requirements of this title.''}\footnote{%
Patent laws can be found in http://www.uspto.gov/web/offices/pac/mpep/mpep-9015-appx-l.html\#d0e302376} In practice however, other types of invention including algorithms can also be patented.\footnote{A notable example is the patent \textit{US6285999} protecting the Page Rank algorithm invented by Larry Page in 1998 which was the genesis of Google.} The two following sections of the 35 USC defined the condition an invention must meet to be protected by the USPTO: (i) novelty: the claimed invention cannot be already patented or described in a previous publication (35 USC section 102); (ii) obviousness: \textit{``differences between the claimed invention and the prior art must not be such that the claimed invention as a whole would have been obvious before the effective filing date of the claimed invention to a person having ordinary skill in the art to which the claimed invention pertains''}. (35 USC section 103). After review from the USPTO experts, an application satisfying these requirements will be accepted and a patent granted. The average time lag for such a review is on average a little more than 2 years since 1976, with some patents being granted after much more than two years.\footnote{This time lag, sometimes called the grant lag, is highly heterogeneous across technological fields. In addition, it cannot be considered as totally random. For example, if the patent is really disruptive some competitors might have some incentive in delaying the process by disputing the validity of the patent, for more details see~\cite{regibeau2010}.}

\paragraph*{Sample restriction}
As explained briefly before, we consider every patent granted by the USPTO between 1976 and 2013. For each patent, we gather information on the year of application, the year the patent was granted, the name of the inventors, the name of the assignees and the technological fields in which the patent has been classified (we get back to what these fields are below). We restrict attention to patents applied for before 2007. The choice of the year 2007 is due to the truncation bias: we only want to use information on granted patents and we get rid of all patents that were rejected by the USPTO. However, in order to date them as closely as possible to the date of invention, we use the application date as a reference. As a consequence, as we approach the end of the sample, we only observe a fraction of the patents which have been granted by 2013. Looking at the distribution of time lag between application and grant in the past and assuming that this distribution is complete in time, we can consider that data prior to 2007 are almost complete and that data for 2007 are complete up to 90\%.

\newpage

\section*{S2 Text : Data collection and Workflow}

\label{supp:data}

\subsubsection*{Data Collection Procedure}

Raw version of USPTO redbook with abstracts are available for years 1976-2014 starting from bulk download page at \url{https://bulkdata.uspto.gov/}. A script first automatically downloads files. Before being automatically processed, a few error in files (corresponding to missing end of records probably due to line dropping during the concatenation of weekly files) had to be corrected manually. Files are then processed with the following filters transforming different format and xml schemes into a uniform dictionary data structure :

\begin{itemize}
\item \texttt{dat} files (1976-2000): handmade parser
\item \texttt{xml} files (2001-2012): xml parser, used with different schemas definitions.
\end{itemize}

Everything is stored into a MongoDB database, which latest dump is available at 
\url{http://dx.doi.org/10.7910/DVN/BW3ACK} 

\subsubsection*{Processing Workflow}

The source code for the full workflow is available at \url{https://github.com/JusteRaimbault/PatentsMining}. A simplified shell wrapper is at \texttt{Models/fullpipe.sh}. Note that keywords co-occurrence estimation requires a memory amount in $O(N^2)$ (although optimized using dictionaries) and the operation on the full database requires a consequent infrastructure. Launch specifications are the following :

\paragraph*{Setup} 

Install the database and required packages.

\begin{itemize}
\item Having a running local mongod instance
\item mongo host, port, user and password to be configured in \texttt{conf/parameters.csv}
\item raw data import from gz file : use mongorestore -d redbook -c raw --gzip {\$}FILE
\item specific python packages required : pymongo, python-igraph, nltk (with resources punkt, averaged{\_}perceptron{\_}tagger,porter{\_}test)
\end{itemize}

\paragraph*{Running}

The utility fullpipe.sh launches the successive stages of the processing pipe.

\paragraph*{Options}

\textit{this configuration options can be changed in }\texttt{conf/parameters.csv}

\begin{itemize}
\item window size in years
\item beginning of first window
\item beginning of last window
\item number of parallel runs
\item \texttt{kwLimit} : total number of keywords $K_W$
\item \texttt{edge{\_}th} : $\theta_w$ pre-filtering for memory storage purposes
\item \texttt{dispth} : $\theta_c$
\item \texttt{ethunit} : $\theta_w^{(0)}$
\end{itemize}

\paragraph*{Tasks}

The tasks to be done in order : keywords extraction, relevance estimation, network construction, semantic probas construction, are launched with the following options :

\begin{enumerate}
\item \texttt{keywords} : extracts keywords
\item \texttt{kw-consolidation} : consolidate keywords database (techno disp measure)
\item \texttt{raw-network} : estimates relevance, constructs raw network and perform sensitivity analysis
\item \texttt{classification} : classify and compute patent probability, keyword measures and patent measures ; here parameters $(\theta_w,\theta_c)$ can be changed in configuration file.
\end{enumerate}

\paragraph*{Classification Data}

The data resulting from the classification process with parameters used here is available as \texttt{csv} files at\\
 \url{http://37.187.242.99/files/public/classification_window5_kwLimit100000_dispth0.06_ethunit4.1e-05.zip}. Each files are named according to their content (keywords, patent probabilities, patent measures) and the corresponding time window. The format are the following :

\begin{itemize}
\item Keywords files : keyword ; community ; termhood times inverse document frequency ; technological concentration ; document frequency ; termhood ; degree ; weighted degree ; betweenness centrality ; closeness centrality ; eigenvector centrality
\item Patents measures : patent id ; total number of potential keywords ; number of classified keywords ; same topological measures as for keywords
\item Patent probabilities : patent id ; total number of potential keywords ; id of the semantic class ; number of keywords in this class. Probabilities have to be reconstructed by extracting all the lines corresponding to a patent and dividing each count by the total number of classified keywords.
\end{itemize}

\paragraph*{Analysis}

The results of classification has to be processed for analysis (construction of sparse matrices for efficiency e.g.), following the steps:
\begin{itemize}
\item from classification files to R variables with \texttt{Semantic/semanalfun.R}
\item from csv technological classes to R-formatted sparse Matrix with \texttt{Techno/prepareData.R} 
\item from csv citation file to citation network in R-formatted graph and adjacency sparse matrix with \texttt{Citation/constructNW.R}
\end{itemize}

Analyses are done in \texttt{Semantic/semanalysis.R}.

\newpage

\section*{S4 Text : Network Sensitivity Analysis \label{app:sensitivity}}

\subsubsection*{Network Sensitivity}

\begin{figure}
\centering
\includegraphics[width=\textheight,height=\textwidth,angle=90]{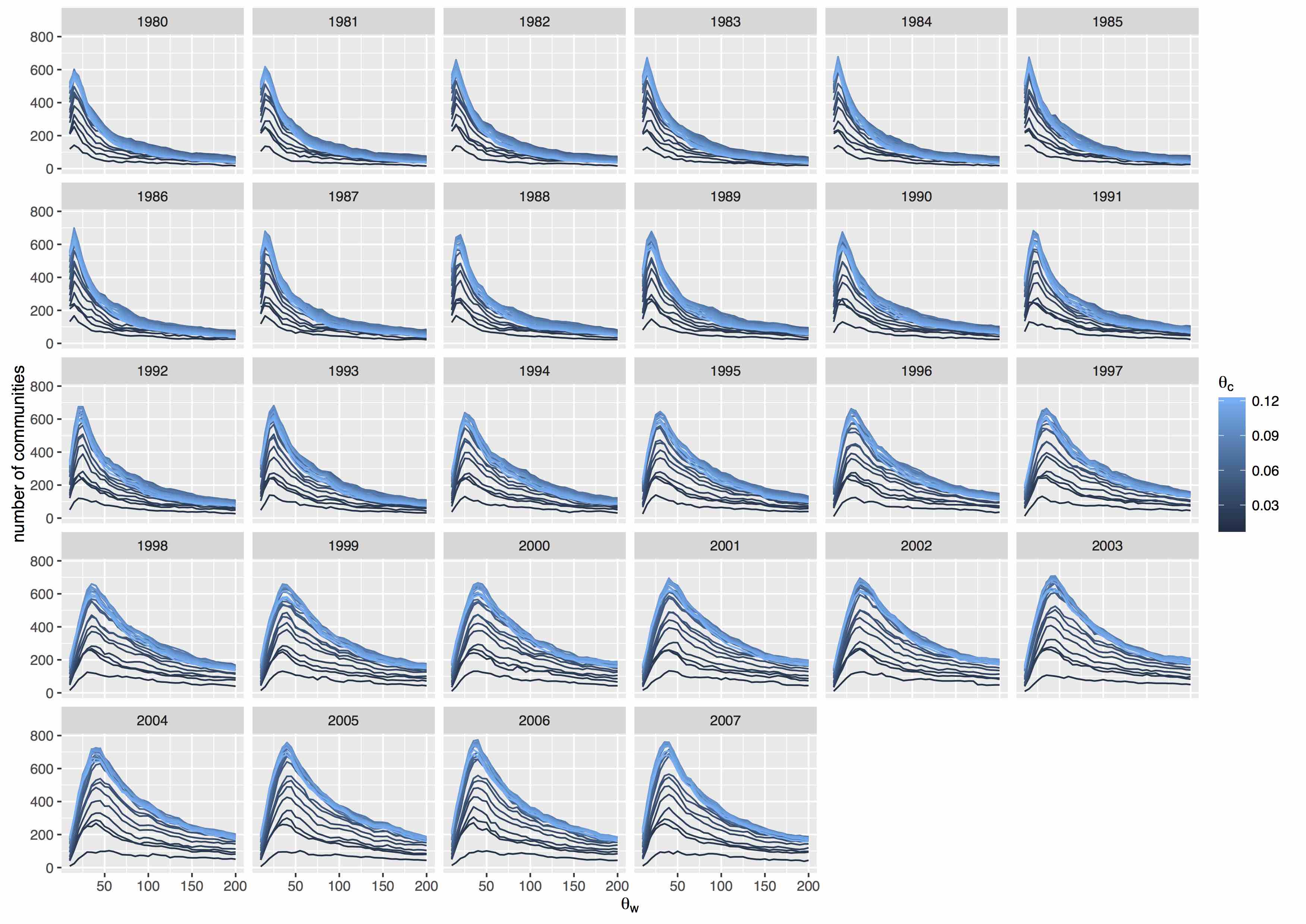}
\caption{Sensitivity plots for $T_0 = 4$ : Number of communities as a function of $\theta_w$, for each year.}
\label{fig:ext-sensitivity-1}
\end{figure}

\begin{figure}
\centering
\includegraphics[width=\textheight,height=\textwidth,angle=90]{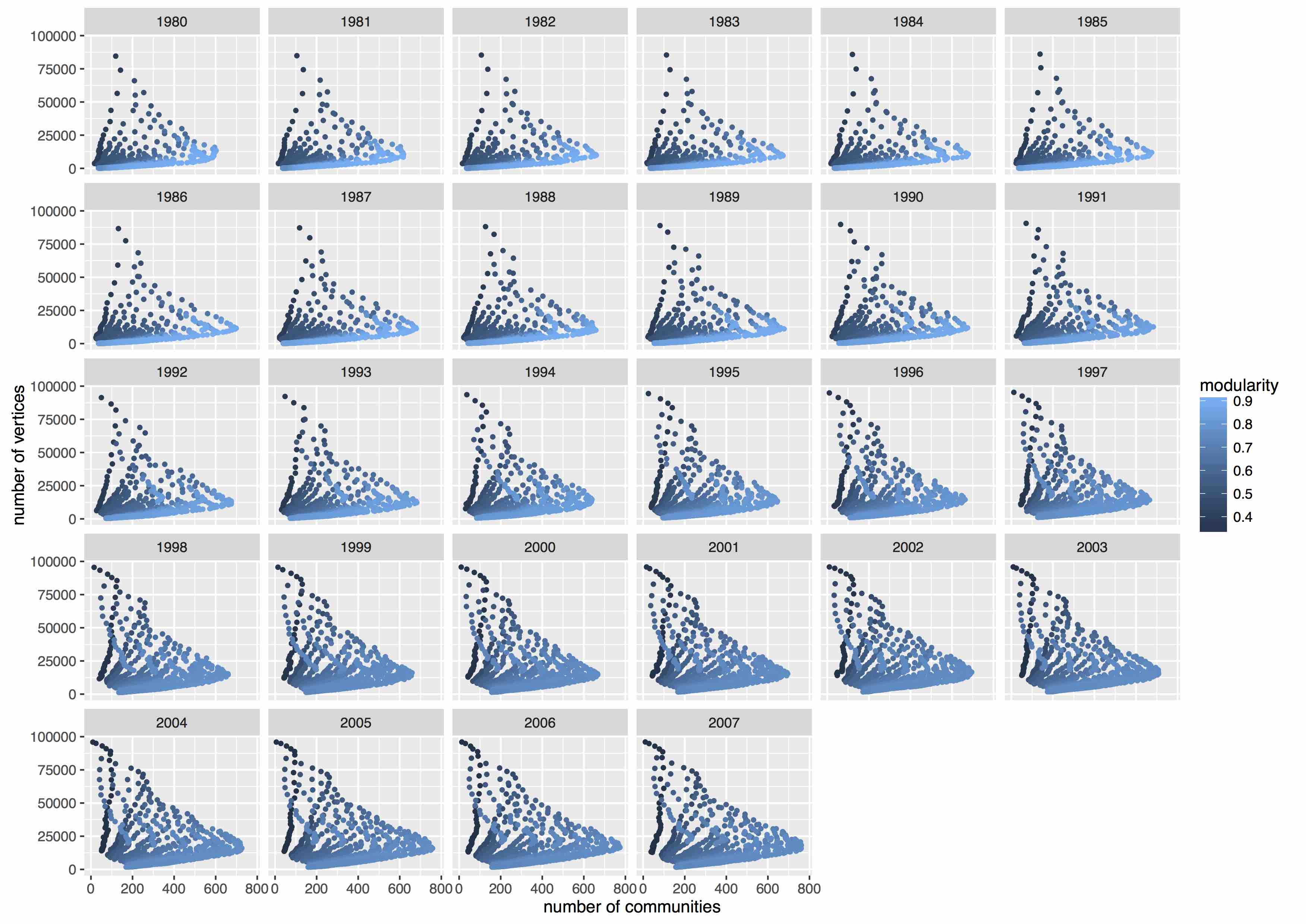}
\caption{Sensitivity plots for $T_0 = 4$ : Pareto plots of number of communities and number of vertices, for each year.}
\label{fig:ext-sensitivity-2}
\end{figure}

\begin{figure}
\centering
\includegraphics[width=\textheight,height=\textwidth,angle=90]{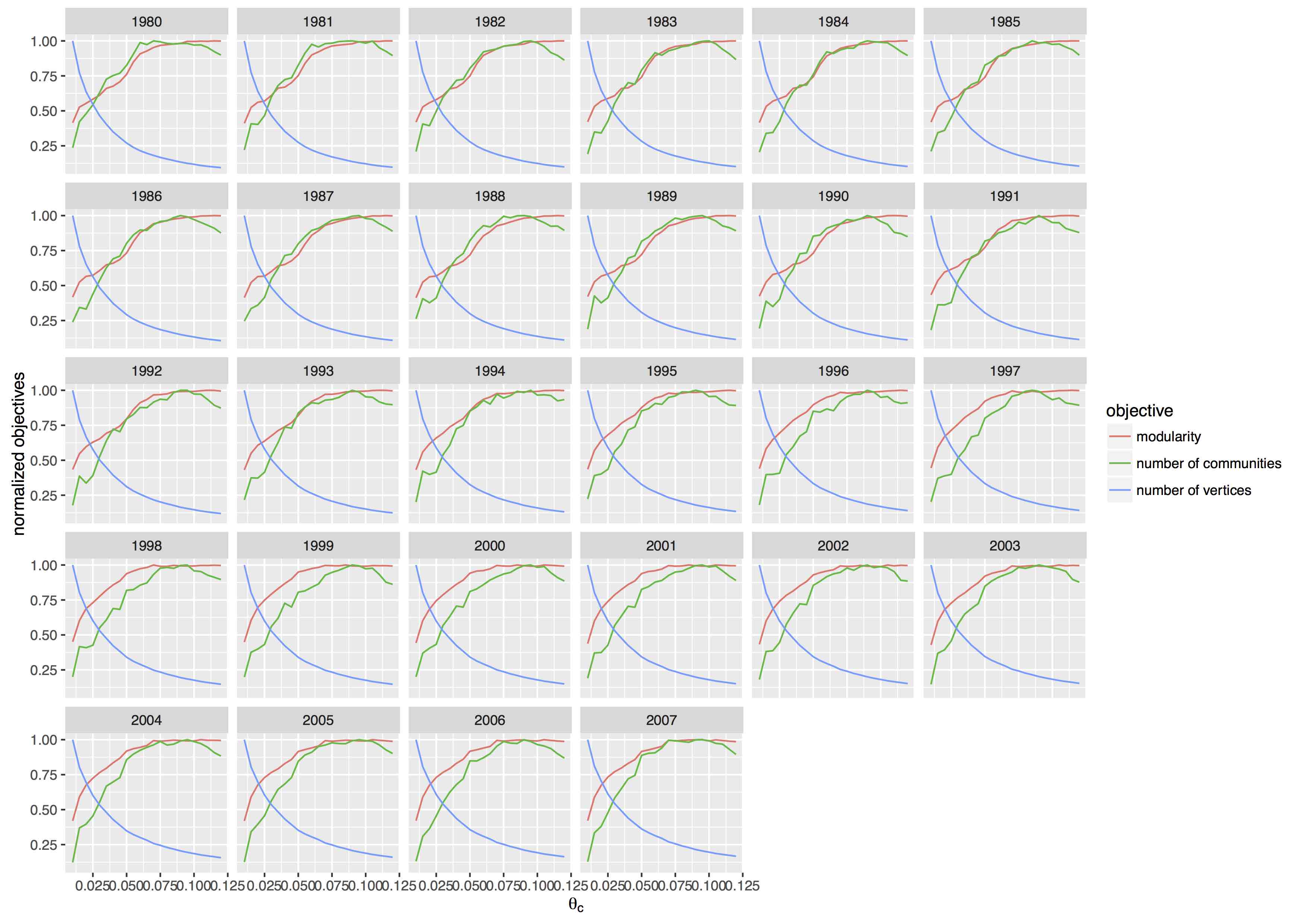}
\caption{Sensitivity plots for $T_0 = 4$ : normalized objective as a function of $\theta_c$, for each year.}
\label{fig:ext-sensitivity-3}
\end{figure}

The example of Fig.1 in main text for a given year yielded the same qualitative behavior for all years, as shown in Fig.~\ref{fig:ext-sensitivity-1}, \ref{fig:ext-sensitivity-2} and \ref{fig:ext-sensitivity-3} here.  We also show an other point of view over the Pareto optimization, that is the third plot giving the values of normalized objectives as a function of $\theta_c$.


\subsubsection*{Time-window size sensitivity}

\begin{figure}
\centering
\includegraphics[width=\textheight,height=\textwidth,angle=90]{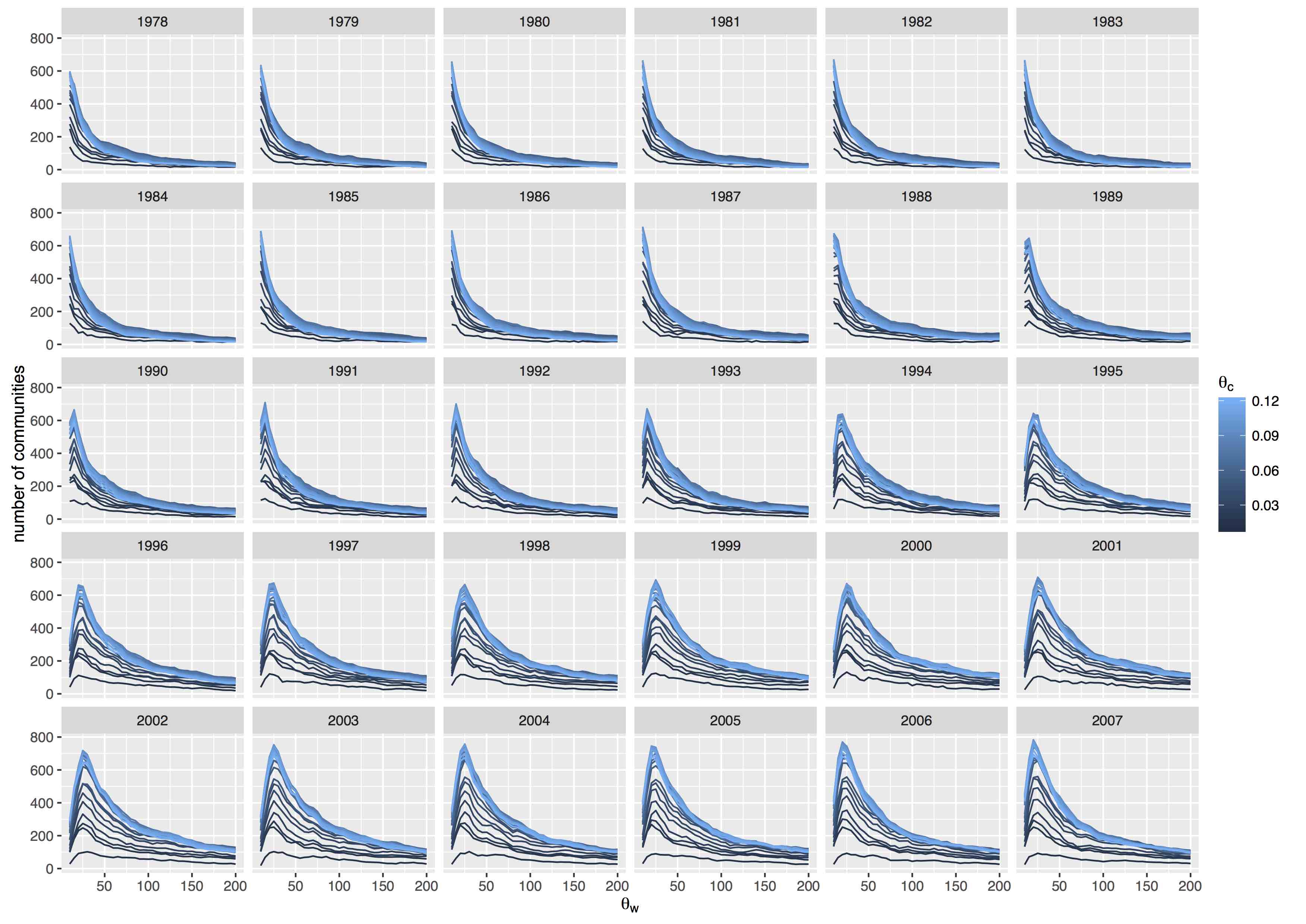}
\caption{Sensitivity plots for $T_0 = 2$ : Number of communities as a function of $\theta_w$, for each year.}
\label{fig:sensitivity-window3-1}
\end{figure}
\begin{figure}
\centering
\includegraphics[width=\textheight,height=\textwidth,angle=90]{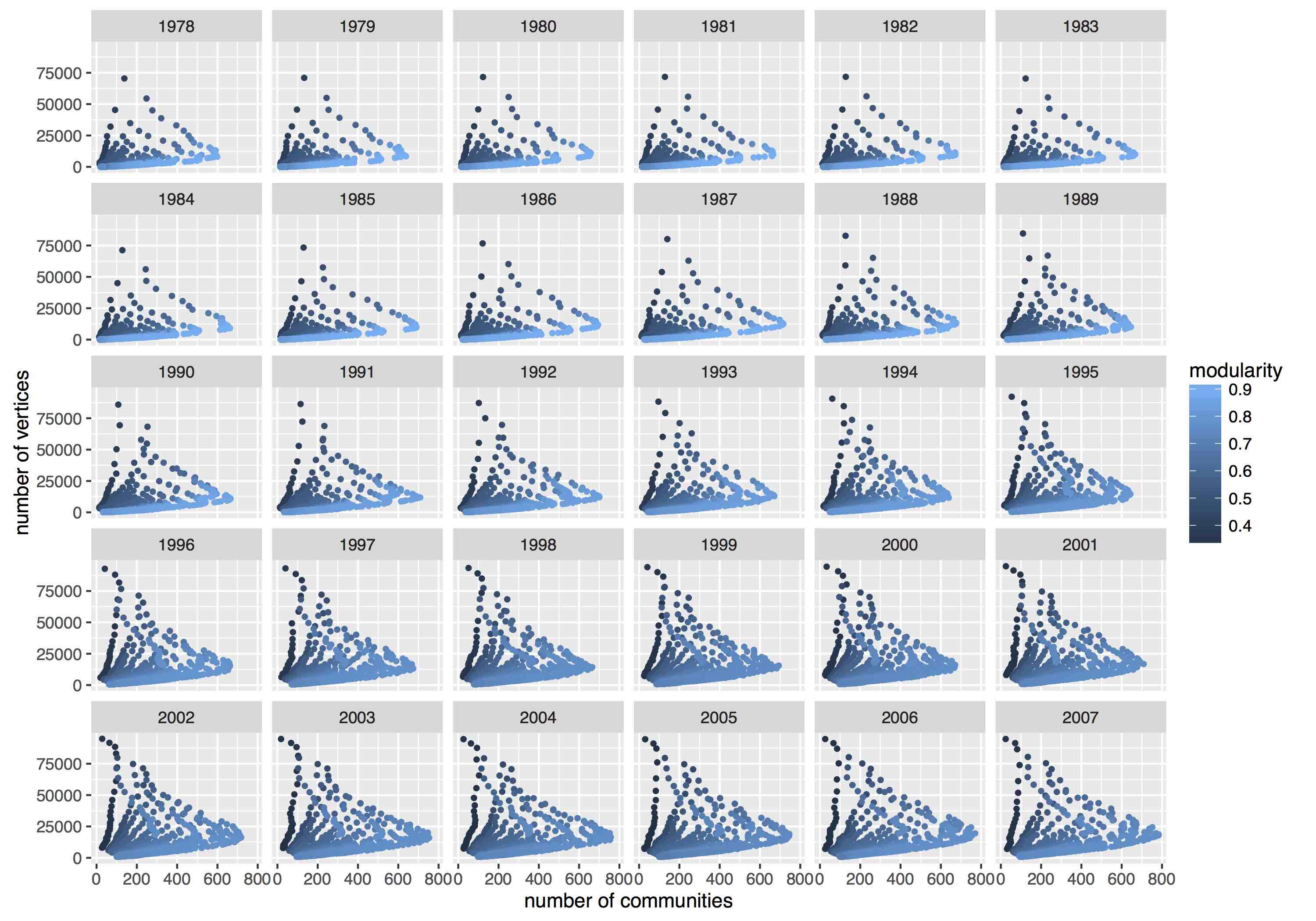}
\caption{Sensitivity plots for $T_0 = 2$ : Pareto plots of number of communities and number of vertices, for each year.}
\label{fig:sensitivity-window3-2}
\end{figure}
\begin{figure}
\centering
\includegraphics[width=\textheight,height=\textwidth,angle=90]{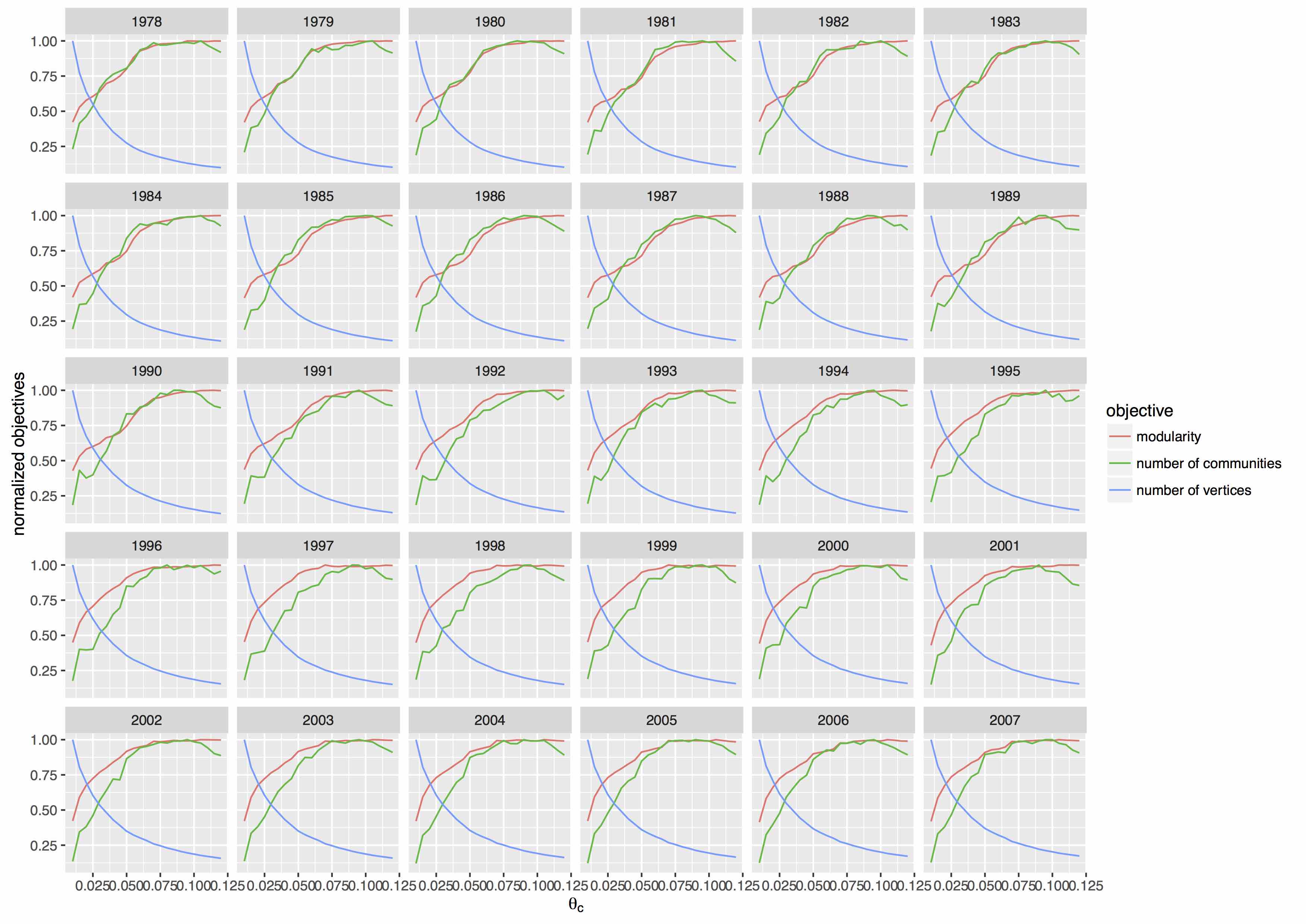}
\caption{Sensitivity plots for $T_0 = 2$ : normalized objective as a function of $\theta_c$, for each year.}
\label{fig:sensitivity-window3-3}
\end{figure}

We show in Fig.~\ref{fig:sensitivity-window3-1}, \ref{fig:sensitivity-window3-2} and \ref{fig:sensitivity-window3-3} the sensitivity plots used for semantic network construction optimization, for a different time window with $T_0 = 2$. The same qualitative behavior is observed (with different quantitative values, as typically $\theta_w^{(0)}$ is for example expected to vary with document number and semantic regime, thus with window size), what confirms that the method is valid across different time windows.

\newpage

\section*{S5 Text : Statistical definitions and derivations}
\subsubsection*{Likelihood expression}
We define $\mathcal{F}_i$ the filtration which corresponds to the time $t_i$. With this notation $\mathcal{L}( X | \mathcal{F}_{i-1})$ simply means the likelihood of $X$ conditioned on the past. We consider $\widehat{\theta}^{(z)}$ the MLE\footnote{Apparently, this MLE is a partial MLE, but we will not refer to partial for simplicity.} of $\theta^{(z)}$, where the corresponding log-likelihood of the model can be expressed up to constant terms as
\begin{eqnarray*}
\sum_{i=2}^Z \sum_{l=1}^{C_i} \log \mathcal{L} \big( B_{l,i} | \mathcal{F}_{i-1} \big).
\end{eqnarray*}
Recalling that $B_{l,i}$ are independent of each other and conditioned on the past follow Bernoulli variables 
$$B \Big( \min \Big\{ 1, \frac{N_{i}^{(z)}}{i-1} + \theta^{(z)} \Big\} \Big),$$ 
the log-likelihood of the model can be expressed as 
\begin{eqnarray}
\label{loglik}
\sum_{i=2}^Z \sum_{l=1}^{C_i} B_{l,i} \log \Big( \min \Big\{ 1, \frac{N_{i}^{(z)}}{i-1} + \theta^{(z)} \Big\} \Big) +(1 - B_{l,i}) \log \Big(1 - \min \Big\{ 1, \frac{N_{i}^{(z)}}{i-1} + \theta^{(z)} \Big\} \Big).
\end{eqnarray}
In practice, the user can easily implement the formula (\ref{loglik}) for any $0 \leq \theta^{(z)} \leq 1$, and maximize it over a predefined grid to obtain $\widehat{\theta}^{(z)}$.

\subsubsection*{Standard errors of the estimated values}
Under some assumptions, it is possible to show the asymptotic normality of $\widehat{\theta}^{(z)}$ and to compute the asymptotic variance. For simplicity of exposition, we assume that we restrict to $\theta^{(z)}$ such that we have $\frac{N_{i}^{(z)}}{i-1} + \theta^{(z)} < 1$ for any $i=2, \cdots, Z$. The central limit theorem can be expressed as 
\begin{eqnarray}
\label{variance}
\sqrt{Z \mathbb{E} [ C ]} (\widehat{\theta}^{(z)} - \theta^{(z)}) \overset{\mathcal{L}}{\rightarrow} MN \Big( 0, \int (p + \theta^{(z)})(1 - (p + \theta^{(z)})) d\pi^{(z)} (p) \Big),
\end{eqnarray}
where MN stands for a multinormal distribution and $\pi^{(z)}$ for the asymptotic limit distribution of the quantity $\frac{N_{i}^{(z)}}{i-1} + \theta^{(z)}$. Note that the variance term in (\ref{variance}) is equal to an aggregate version of the Fisher information matrix. The proof of such statement is beyond the scope of this paper. On the basis of (\ref{variance}), we provide a variance estimator as 
$$v^{(z)} = \frac{1}{C_k-1}\sum_{i=2}^{C_k}  \frac{N_{i_k}^{(z)}}{i-1} + \widehat{\theta}^{(z)},$$ 
where $i_k$ is such that the $i_k$th patent corresponds to the $k$th couple. This estimator was used to compute the standard deviation in Table \ref{summary}.
\subsubsection*{Test statistic}
The test statistic used is a mean difference test statistic between $\widehat{\theta}^{(tec)}$ and  $\widehat{\theta}^{(sem)}$, where the formal expression can be found in (\ref{teststat}). We assume independence between both quantities and thus under the null hypothesis, we have that 
$$\widehat{\theta}^{(tec)} - \widehat{\theta}^{(sem)} {\rightarrow} MN(0,V),$$
where $V= \int (p + \theta^{(tec)})(1 - (p + \theta^{(tec)})) d\pi^{(tec)} (p) + \int (p + \theta^{(sem)})(1 - (p + \theta^{(sem)})) d\pi^{(sem)} (p)$ can be estimated by $\widehat{V} = v^{(sem)} + v^{(tec)}$. Then, we obtain that 
\begin{eqnarray}
\label{teststat}
A = \frac{\widehat{\theta}^{(tec)} - \widehat{\theta}^{(sem)}}{\widehat{V}} \approx \mathcal{N} (0,1),
\end{eqnarray}
where $A$ is the mean difference test static.

\end{document}